\documentclass{amsart}
\usepackage{amsthm, amsmath, amssymb, enumerate}
\usepackage{float}
\usepackage{amsmath}
\usepackage{graphicx}
\usepackage{dcolumn}
\usepackage{bm}
\usepackage[T1]{fontenc}
\usepackage{textcomp}
\usepackage{tikz}
\usepackage{amssymb}
\usepackage{soul}
\usepackage{bbold}
\usepackage{xcolor}
\usetikzlibrary{positioning,arrows}
\usetikzlibrary{decorations.pathmorphing}
\usetikzlibrary{decorations.markings}
\definecolor{myblue}{RGB}{56,94,141}

\newcommand{\newc}{\newcommand}

\newc{\kt}{\rangle}

\newc{\pr}{\prime}
\newc{\longra}{\longrightarrow}
\newc{\ot}{\otimes}
\newc{\rarrow}{\rightarrow}
\newc{\h}{\hat}
\newc{\bom}{\boldmath}
\newc{\btd}{\bigtriangledown}
\newc{\al}{\alpha}
\newc{\be}{\beta}
\newc{\ld}{\lambda}
\newc{\sg}{\sigma}
\newc{\p}{\psi}
\newc{\eps}{\epsilon}
\newc{\om}{\omega}
\newc{\mb}{\mbox}
\newc{\tm}{\times}
\newc{\hu}{\hat{u}}
\newc{\hv}{\hat{v}}
\newc{\hk}{\hat{K}}
\newc{\ra}{\rightarrow}
\newc{\non}{\nonumber}

\newc{\dg}{\dagger}
\newc{\prh}{\mbox{PR}_H}
\newc{\prq}{\mbox{PR}_q}
\newc{\pd}{\partial}
\newc{\qv}{\vec{q}}
\newc{\pv}{\vec{p}}
\newc{\dqv}{\delta\vec{q}}
\newc{\dpv}{\delta\vec{p}}
\newc{\mbq}{\mathbf{q}}
\newc{\mbqp}{\mathbf{q'}}
\newc{\mbpp}{\mathbf{p'}}
\newc{\mbp}{\mathbf{p}}
\newc{\mbn}{\mathbf{\nabla}}
\newc{\dmbq}{\delta \mbq}
\newc{\dmbp}{\delta \mbp}
\newc{\T}{\mathsf{T}}
\newc{\J}{\mathsf{J}}
\newc{\sfL}{\mathsf{L}}
\newc{\C}{\mathsf{C}}
\newc{\B}{\mathsf{M}}
\newc{\V}{\mathsf{V}}
\usepackage{titlesec}

\titleformat{\title}{\centering\LARGE\bfseries}{\thetitle}{1em}{}
\titleformat{\section}{\centering\LARGE\bfseries}{\thesection}{1em}{}
\titleformat{\subsection}{\Large\bfseries}{\thesubsection}{1em}{}

\begin{document}

\title{Hybrid quantum-classical chaotic NEMS}

\author{A. K. Singh}
\address{Department of Physics, Indian Institute of Technology (Banaras Hindu University) Varanasi - 221005, India}
\author{L.~Chotorlishvili}
\address{Department of Physics and Medical Engineering, Rzeszow University of Technology, 35-959 Rzeszow, Poland}
\author{Z.~Toklikishvili}
\address{Faculty of Mathematics and Natural Sciences, Tbilisi State University, Chavchavadze av.3, 0128 Tbilisi, Georgia}
\author{I.~Tralle}
\address{Department of Physics and Medical Engineering, Rzeszow University of Technology, 35-959 Rzeszow, Poland}
\author{S. K. Mishra}
\address{Department of Physics, Indian Institute of Technology (Banaras Hindu University) Varanasi - 221005, India}
\email{Corresponding author:abhishekkrsingh.rs.phy17@itbhu.ac.in }

\begin{abstract}
We present an exactly solvable model of a hybrid  quantum-classical system of a Nitrogen-Vacancy (NV) center spin (quantum spin) coupled to a nanocantilever (classical) and analyze the enforcement of the regular or chaotic classical dynamics onto the quantum spin dynamics.
The main problem we focus in this paper is whether the classical dynamical chaos may induce chaotic effects in the quantum spin dynamics or not. We explore several characteristic criteria of the quantum chaos, such as quantum Poincar\'{e} recurrences, generation of coherence and energy level distribution and observe interesting chaotic effects in the spin dynamics. Dynamical chaos imposed in the cantilever dynamics through the kicking pulses induces the stochastic dynamics on the quantum subsystem.  We consider a quantum system of two and three levels and show that in a two-level case, type of stochasticity is not conforming all the characteristic features of the quantum chaos and is distinct from it. We also explore the effect of quantum feedback on dynamics of the cantilever and the entire system.
\end{abstract}

\date{\today}

\maketitle


\section{Introduction}
Classical or quantum finite systems may show non-deterministic behaviour when coupled to a stochastic bath (or other external randomness sources), or nonlinearity \cite{el2007deterministic,PhysRevA.100.062107}. In such hybrid systems any small perturbation destroys the regular motion and leads to unpredictable evolution of the system.  In the first case, the stochasticity appears to be external, and in the second case, it is an intrinsic property of the system \cite{cresson2019selection}. For example, in the case of a kicked rotator in classical regime, when the strength of kicking is increased, the regular periodic motion is destroyed and chaotic motion is observed. The chaotic behaviour can be validated by a diffusive growth in the kinetic energy \cite{chirikov1979universal,casati1979stochastic} of the kicked rotator. The quantum delta kicked rotators play an important role in understanding quantum chaos and other related effects \cite{lichtenberg2013regular}. The existence of quantum resonance can be seen using the quantum kicked rotators \cite{moore1995atom}. Experimentally the quantum kicked rotators and quantum chaos can be studied using ultracold atoms which are driven by periodically kicked by optical pulses \cite{PhysRevLett.87.074102}. The effect of the nonlinearity on a two-level system coupled with kicked rotor is already studied \cite{Tanaka_1996}.
Dynamical chaos refers to the phenomenon of extreme sensitivity of phase trajectories to a tiny disturbance. It is worth to note that in the quantum case, we do not have phase trajectories and the chaos is manifested in the Gaussian statistics of the energy spectrum
\cite{casati1985energy}.  The remarkable feature of quantum chaos is the termination of classically allowed diffusive processes leading to destruction of quantum coherence
\cite{shepelyansky1987localization,PhysRevA.44.4704}. In this paper our interest is in a hybrid system under the constraint such that part of the system is classical, and the rest is quantum. In such a quantum-classical hybrid system  when the classical part exhibits dynamical chaos, we analyze the spread of chaos to the quantum part. 

 In the last few years the hybrid systems consisting of the spin and mechanical parts named as Nano-electromechanical systems (NEMS) generated a lot of interest  \cite{Naik2009,Connell2010,Alegre2011,Stannigel2010,Safavi-Naeini2011,Camerer2011,Eichenfield2009,Safavi-Naeini2012,Brahms2012,Nunnenkamp2012,Khalili2012,Meaney2011,Atalaya2011,Rabl2010,
Prants2011,Ludwig2010,Schmidt2010,Karabalin2009,Chotorlishvili_2011,
Shevchenko2012,Liu2010,Shevchenko2010,Zueco2009,Cohen2013,Rabl2009,
Zhou2010,Chotorlishvili2013}.  In these systems the spin subsystem is always described quantum-mechanically and the
mechanical subsystem ( {\it i. e.}, the oscillator) can be considered either in quantum or classical (linear or nonlinear) regimes. All these cases need special mathematical description and show the realization of a physical
features. Our interest here concerns the case when cantilever coupled to the NV center performs nonlinear oscillations. For more details about the model in question, we refer to the earlier works \cite{Chotorlishvili_2011,Chotorlishvili2013,Toklikishvili2017}.
In particular, we assume that kicks of the external driving field force the classical motion of the cantilever and the overlapping of nonlinear resonances may induce the chaotic motion of the cantilever. The chaoticity of the motion of the cantilever extends to the quantum spin dynamics via the cantilever-spin coupling term.
 The spin of the NV center is described by spin triplet $S=1$, with $m_s=-1,\ 0,\ {\rm{and}} \ 1$. States $\vert-1\rangle$ and $\vert 1\rangle$ are separated by potential barrier $D\hat S_z^2\approx \hbar\omega_0$, where $\omega_0=2.88$GHz. In what follows we set $\hbar=1$. Hamiltonian of the NV center has the form \cite{Rabl2009}:
\begin{eqnarray}\label{general NV form}
\hat H_{NV}=\sum\limits_{i=\pm 1}\left(-\delta_i\vert i\rangle\langle i\vert+\frac{\Omega_i}{2}(\vert 0\rangle\langle i\vert+\vert i\rangle\langle 0\vert)\right),
\end{eqnarray}
where $\delta_i$ and $\Omega_i$ are detunings and Rabi frequencies of the two microwave (MW) transitions. In the limit of single MW field and zero external magnetic field $B_z\rightarrow 0$, Hamiltonian Eq.(\ref{general NV form}) couples the ground state $\vert 0\rangle$ with bright superposition of the excited states $\vert b\rangle=(\vert -1\rangle+\vert 1\rangle)/\sqrt{2}$, while dark state $\vert d\rangle=(\vert -1\rangle-\vert 1\rangle)/\sqrt{2}$ is decoupled and NV center reduces to an effective two-level model. In what follows we consider both two- and three-level problems.
\begin{figure}[t!]
 \includegraphics[width=0.45\linewidth,height=2.2in]{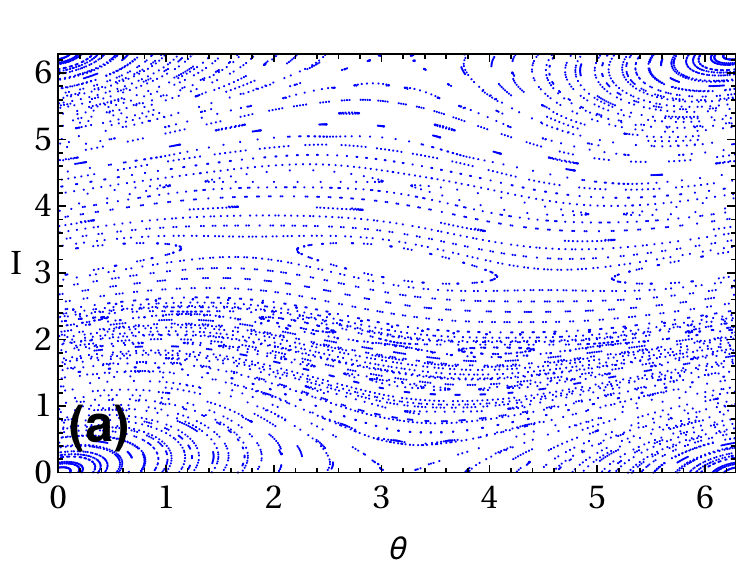}\ \includegraphics[width=0.45\linewidth,height=2.2in]{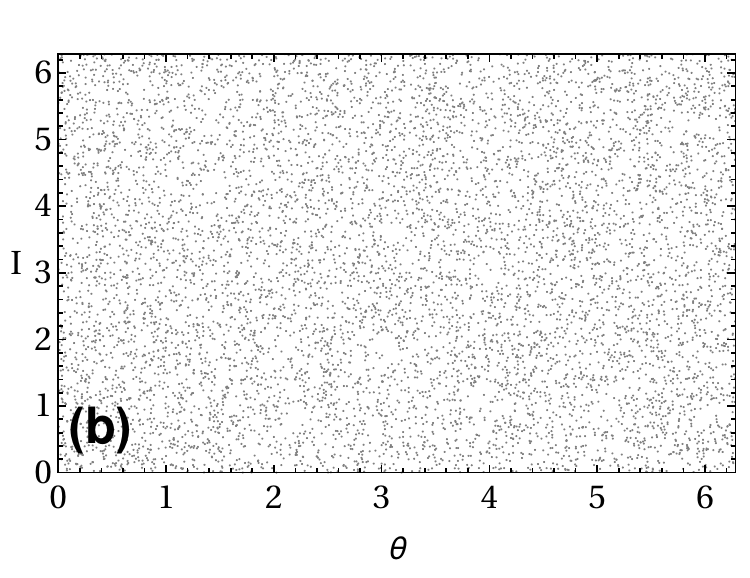}
 \caption{The Phase space plot of cantilever's dynamics constructed through the recurrence relations Eq. (\ref{recurrence relation}) in (a) the regular regime $K=0.5$ (Blue) where the phase space is covered by two different phase trajectories: open hyperbolic and some part of closed elliptic, and (b) the chaotic regime at $K=10$ (Gray) where the entire phase space is covered by a chaotic sea. Topologically different phase trajectories are bordered by separatrix line. The values of parameters are: $K=\epsilon I_{0} T \frac{6\pi \mu}{m^2\omega_{r}^{2}}$, $\mu=\frac{\omega_{r}^{2}m}{2a_{0}^2}$ $I_{0}=\frac{m}{2}x_{0}^2 \omega_{r}$, $m=6\times10^{-17}$Kg, $x_{0}=a_{0}=5\times10^{-3}$m, $T=10\mu$s, $\omega_{r}=\omega_{0}=2\pi\times5\times10^{6}$Hz, for chaotic case $\varepsilon=0.003$ and for the regular case $\varepsilon=0.0003$ .}
\label{regular}
 \end{figure}
In the case of coupling with the bright state the Hamiltonian of the hybrid system of NV center and a driven nonlinear oscillator is given as \cite{Zhou2010}
\begin{eqnarray}\label{Lorentz}
H\big(x,p,t\big)=H_{S}+H_{0}\big(x,p\big)+H_{NL}
+\varepsilon V\big(x,t\big)+g \hat{V}_{c,NV}.
\end{eqnarray}
Here $\hat{H}_{S}=\frac{1}{2}\omega_{0}\hat{\sigma}_{z}$ is the Hamiltonian of the NV center, The splitting frequency is given by $\omega_{0}=\big(\omega_{R}^{2}+\delta^{2}\big)^{1/2}$, where $\omega_{R}$ is the Rabi frequency, and $\delta$ is the detuning. The spin operator $\hat{S}_{z,NV}$ can be written in the basis of NV center  as \cite{Zhou2010}:
$\hat{S}_{z,NV}=\frac{1}{2}\big(\cos(\alpha)\hat{\sigma}_{z}+\sin(\alpha)(\hat{\sigma}_{+}+\hat{\sigma}_{-})\big)$ with $\tan(\alpha)=-\omega_{R}/\delta$ and $\hat{\sigma}_{z}=|e\rangle\langle e|-|g\rangle\langle g|$, $\hat{\sigma}_{+}=|e\rangle\langle g|,~~\hat{\sigma}_{-}=|g\rangle\langle e|$.
The linear part of the oscillator is given by term $H_{0}=\frac{p^{2}}{2m}+\frac{\omega_{r}^{2}mx^{2}}{2}$ and the nonlinear part is given by $H_{NL}=\beta x^{3}+\mu x^{4}$,
where $\omega_{r}$ is the frequency of the oscillations, $\beta$ and $\mu$ are constants of the nonlinear terms.
The term
\begin{eqnarray}
&&V\big(x,t\big)=V_{0}xT\sum\limits_{n=-\infty}^{\infty}\delta\left(t-nT\right),\nonumber\\
&&\varepsilon V_{0}=f_{0},~~\varepsilon \ll 1,
\label{hamiltonian2}
\end{eqnarray}
describes the driven motion of the cantilever in the microwave field of delta pulses with frequency $\omega=2\pi/T$.
The key issue is the last term $\hat{V}_{c,NV}=x(t)\hat{S}_{z,NV}$ in Eq.~(\ref{Lorentz}) which describes the coupling between the classical cantilever and the quantum NV spin. The distance and the coupling strength between the magnetic tip and NV spin depend on the magnetostriction effect \cite{Chotorlishvili2013}. Subject to the cantilever's oscillations, $x(t)$ can be either chaotic or regular. In what follows, we show that classical dynamical chaos leads to the stochastic phenomenon in spin dynamics. 
We consider two and three level models of the NV center and show that in both cases chaotic dynamics of the cantilever leads to the chaotic spin dynamics. However, being stochastic, two-level model does not manifest all the characteristic features of quantum chaos.
In the present manuscript our main focus is a mathematical formulation of the problem. Nevertheless, we specify the values of the parameters relevant to the NV centers \cite{Rabl2009}: $\frac{\omega_{r}}{2\pi}=5$ MHz, $\frac{\omega_{R}}{2\pi}=0.1-10$  MHz, $\delta=1$ kHz, mass of the cantilever $m=6\times 10^{-17}$ kg, the coupling constant $\frac{g}{2\pi}=100$ kHz, the amplitude of the zero point fluctuations $a_{0}=\sqrt{\hbar/2m\omega_{r}}\approx 5\times 10^{-3}$ m. The nonlinear constants are order of $\beta\approx\frac{\omega_{r}^{2}m}{2a_{0}}$, $\mu\approx\frac{\omega_{r}^{2}m}{2a_{0}^{2}}$. The energy scale of the problem is defined by
$\varepsilon V\approx \omega_{r}^{2}ma_{0}^{2}\approx 10^{-9}$J, and the time scale is of order of microsecond scale $t\approx \frac{\pi}{2g}$ microseconds.
In what follows, we explore the spreading of classical dynamical chaos on the quantum system. In the quantum part of the NEMS, spin dynamics manifest some characteristic features of the quantum chaos \cite{Chotorlishvili2010,zaslavsky2007physics, Chotorlishvili2018,ugulava2005irreversible}, but not all of them. Therefore, we term this phenomenon as a hybrid quantum-classical chaos.  The work is organized as follows: In section \ref{cantilever}, we discuss the classical chaotic dynamics of the cantilever. In section \ref{spin_dynamics}, we present analytical results for spin dynamics of NV center spin attached to the cantilever and discuss different aspects of the quantum chaos, namely, quantum coherence, Poincar\'{e} recurrences and level statistics. Subsequently in section \ref{SPIN1} we study dynamics of a three-level NV system. Later, in section \ref{StatisticalSPIN1/2} we explore statistical average over various $I_{0}$ and $\theta_{0}$. In section \ref{Feedback Effect} we study about feedback effect and finally summarize the manuscript in section \ref{summary}.

\section{Dynamics of the cantilever}
\label{cantilever}
The experimentally feasible NEMS consists of the spin of the NV center interacting with a magnetic tip (attached to the end of the nano-cantilever).  The oscillations performed by the cantilever can be viewed as classical or quantum, depending on the simple criteria: At temperatures $T<<2\pi\hbar\omega_r/k_B$, where $k_B$  is the Boltzmann constant and $\omega_r$ is the oscillation frequency, dynamics of a cantilever is quantum, and it exerts quantum feedback effect on a spin dynamics. Typically for $\omega_r=1$kHz, $T<50$nK. At higher temperatures, or when the cantilever is controlled externally by a classical field, dynamics is classical. Large-amplitude nonlinear oscillations are entirely classical. Therefore in what follows, we neglect the quantum feedback effect. 

With the purpose of simplicity, the cantilever part of the Hamiltonian $H_{p,q}=H_{0}+H_{NL}+V(x,t)$ can be rewritten in the action-angle canonical variables through the transformation $\Phi=F+I\theta$:
\begin{eqnarray}\label{L3}
&& d\Phi=pdq+\theta dI+\big(H_{I,\theta}-H_{p,q}\big)dt.
\end{eqnarray}
The canonical equations in the new variables are given as
\begin{eqnarray}\label{L4}
  \frac{dI}{dt}&=-\frac{\partial H_{I,\theta}}{\partial\theta}=-\varepsilon \frac{\partial V(I,\theta)}{\partial \theta}T\sum\limits_{n=-\infty}^{\infty}\delta\left(t-nT\right), \nonumber\\
  \frac{d\theta}{dt}&=\frac{\partial H_{I,\theta}}{\partial I}=\omega(I)+\varepsilon \frac{\partial V(I,\theta)}{\partial I}T\sum\limits_{n=-\infty}^{\infty}\delta\left(t-nT\right).
\end{eqnarray}
The presence of the delta function allows us to introduce the Floquet map $\left(I_{n+1},\theta_{n+1}\right)=\mathcal{F}\left(I_n,\theta_n\right)$ and integrate Eq.~(\ref{L4}) exactly as
\begin{eqnarray}\label{recurrence relation}
&& I_{n+1}=I_n-\varepsilon T\frac{\partial V(I_n,\theta_n)}{\partial \theta_n},\nonumber\\
&& \theta_{n+1}=\theta_n+\omega(I_{n+1})T+\varepsilon T\frac{\partial V(I_n,\theta_n)}{\partial I}.
\end{eqnarray}
From the above equation
we deduce the criteria of the dynamical chaos:
\begin{eqnarray}\label{criteria of the chaos}
K=\varepsilon I_{0}T\left|\frac{d\omega(I)}{dI}\right|,
\end{eqnarray}
Using action-angle variable to transform the cantilever part of the Hamiltonian $H_{p,q}=H_{0}+H_{NL}+V(x,t)$, and taking average
with respect to the fast phase $\theta$, the cubic part of $H_{NL}$ will be zero if we take average with respect to fast phase $\theta$ but the quartic term in $H_{NL}$ will survive, so we obtain, from here $\omega(I)=\partial(H_{0}+H_{NL})/\partial I$, $H_{0}(I)=\omega_{r}I+H_{NL}$, $H_{NL}=3\pi\mu\left(\frac{I}{m\omega_{r}}\right)^{2}$, and $V=V_{0}(I)\cos\theta$, $V_0(I)=V_0\sqrt{2I_0/\omega_r}$.
Using the system specific parameters we define the criterion for dynamical chaos as:
\begin{eqnarray}
K=\varepsilon I_{0}T\Big(\frac{6\pi\mu}{m^2\omega_{r}^2}\Big).
\end{eqnarray}
$K$ is also known as a criterion for overlapping of the resonance. If $K<1$ the  phase trajectories are distinguishable from each other and if $K>1$  phase trajectories correspond to chaotic behaviour \cite{zaslavsky2007physics}.

 After the formation of dynamical chaos, the dynamical description of the problem loses the sense.
The dynamics of the cantilever can be explored using a standard map $I_{n+1}=I_{n}-K \sin{\theta_{n}}$, $\theta_{n+1}=\theta_{n}+I_{n+1}$, where $K$ quantifies the criterion for the dynamical chaos.  For further analytic insights, we utilize the methods of non-equilibrium statistical physics and introduce the distribution function $f(\theta, I)$ such that 
$i\frac{\partial f}{\partial t}=\left(\hat{L}_{0}+\varepsilon \hat{L}_{1}\right)$,
where Liouville operators $\hat{L}_{0}$ and $\hat{L}_{1}$ are defined as follows (see \cite{zaslavsky2007physics} for more details)   $\hat{L}_{0}=-i\omega\frac{\partial}{\partial \theta}$, and $
 \hat{L}_{1}=-i\left(\frac{\partial V}{\partial I}\frac{\partial }{\partial \theta}-\frac{\partial V}{\partial
\theta}\frac{\partial }{\partial I}\right)$.
The critical issue is the difference between the time scales of the slow action $I$ and fast angle $\theta$ variables.  The correlation time scale in the system is defined via $ \langle\langle \theta(t)\theta(0)\rangle\rangle=\exp(-t/\tau_c)$. Typically $\tau_c<T<\tau_D$, where $\tau_D$ is the time spent on the substantial change of the action variable. As we see from Eq. (\ref{L4}),  during the interval between the kicks $ T $, the change of the action variable is small and is proportional to $ \varepsilon $. Our interest here concerns the distribution function for the action variable, and the Fokker-Plank equation averaged over the fast angular variable $F(I)=\langle\langle f(I,\theta)\rangle\rangle_{\theta}$. The general structure of the Fokker-Planck equation  reads \cite{zaslavsky2007physics}:
\begin{eqnarray}\label{the general structure}
&& \frac{\partial F(I)}{\partial t}=-\frac{\partial}{\partial I}(A(I)F(I))+\frac{1}{2}
\frac{\partial^2}{\partial I^2}(B(I)F(I)),
\end{eqnarray}
where $A=\frac{1}{T}\langle\langle \Delta I\rangle\rangle_{\theta},$
and $B=\frac{1}{T}\langle\langle (\Delta I)^2\rangle\rangle_{\theta}$.
After calculating coefficients explicitly, we deduce the kinetic equation as
\begin{eqnarray} \label{kinetic equation}
 \frac{\partial F}{\partial t}&=\frac{1}{2}\frac{\partial }{\partial I}D(I)\frac{\partial F}{\partial I},
 \end{eqnarray}
  where $D(I)=\pi\varepsilon^{2}\sum\limits_{m,p}m^{2}|V_{m}|^{2}\delta(m\omega-p\Omega)$, $\Omega=2\pi/T$, $V_{m}$ is the Fourier component of the interaction term and $m,p$ indices take into account the multiple internal resonances.  The solution of the kinetic equation reads as
\begin{eqnarray}\label{Solution kinetic equation}
   \langle I(t)\rangle &=I_0^2+Dt,~
 D&=\frac{1}{2}\varepsilon V^2 T,
\end{eqnarray}
 where $D$ is the diffusion coefficient.
 The dynamics of the cantilever is chaotic for $K>1$ and otherwise $K<1$ is regular (see
Fig.~\ref{regular}). In the chaotic regime we observe a sea of phase points uniformly distributed over the phase space. In the regular regime, two different phase space trajectories cover the entire space. For the standard map described above, if the parameter $K > K_{c}=0.9716$, the stochastic layers start merging; thus, it will create a domain of chaotic motion that covers whole phase space. As $K$ increases, the islands' size decreases, and only the largest of them can be found in the chaotic sea.
Thus solution for cantilever can be written in the discrete form $x_n
=\sqrt{2I_n/m\omega_r}\cos\theta_n$, $p_n=-\sqrt{2I_n\omega_{r}m}\sin\theta_n$.
\section{Spin-1/2 system attached to the cantilever}
\label{spin_dynamics}
\subsection{Evolved in time wave function} \label{Analytical_spin_dynamics}
Let us consider a hybrid system of Quantum NV spin attached to a classical cantilever whose dynamics is calculated from a standard map. From the Hamiltonian given by Eq. (\ref{Lorentz}) we see that the effect of the cantilever in the NV spin is due the the interaction term $\hat{V}_{c,NV}$.  Therefore, the effective Hamiltonian of the NV center attached to the cantilever can be written as:
\begin{eqnarray}\label{effective Hamiltonian}
\hat{H}_{n}=\frac{1}{2}\omega_{0}\hat{\sigma}_{z}+g\hat{V}_{c,NV},
\end{eqnarray}
where $\hat{V}_{c,NV}=\sqrt{2I_n/m\omega_r}\cos\theta_n\hat{S}_{z,NV}$.
It is important to note here that $I_n$ and $\theta_n$ follow the Floquet map given by Eq. (\ref{recurrence relation}).
By varying the parameter $K$ we change the characteristic of the term $\hat{V}_{c,NV}$
from the regular to the chaotic dynamics of the cantilever and explore the spin dynamics in both cases.
Exploiting the Floquet theory \cite{haake1991quantum} we solve the Schr\"{o}dinger equation analytically and get the state after time $t=NT$ as
\begin{eqnarray}
\label{Schrodinger}
&& \vert \psi(t=NT) \rangle=\mathcal{\hat{U}}^{N}|\psi(0)\rangle,
\end{eqnarray}
where $\mathcal{\hat{U}}^N$ is the time evolution operator evolving the system after $N$ kicks and $|\psi(0)\rangle$ is the initial state of the system. 
The Floquet map $\hat{F}_n$ after the $n^{\rm{th}}$ kick is
 $\mathcal{\hat{F}}_{n}=\exp\left(-i \hat H_{n} T\right)$ and
the evolution operator $\mathcal{\hat{U}}^{N}$ is a time-ordered product of $\mathcal{\hat{F}}_n^s$ given as $\mathcal{\hat{U}}^{N}=\mathcal{\hat{F}}_N\cdots\mathcal{\hat{F}}_{n+1}\mathcal{\hat{F}}_n \mathcal{\hat{F}}_{n-1}\cdots \mathcal{\hat{F}}_3 \mathcal{\hat{F}}_2 \mathcal{\hat{F}}_1$. The exact wave function after time $t=NT$ can be written in the form
\begin{eqnarray}
\label{wave function}
\vert\psi(t=NT)\rangle=\sum\limits_{\lbrace\alpha_n\rbrace=\pm}\left\lbrace\prod\limits_{n=2}^{N}e^{-i\alpha_n\varphi_n}
 \big\langle \varphi_{n}^{\alpha_n}\big\vert\varphi_{n-1}^{\alpha_{n-1}}\big\rangle
 \right\rbrace e^{-i\alpha_1\varphi_1}\big\langle \varphi_{1}^{\alpha_1}\big\vert\psi(0)\big\rangle\big\vert\varphi_{N}^{\alpha_N}\big\rangle.\nonumber\\
\end{eqnarray}
Here  ${\alpha_n}\varphi_{n}$ and $\vert\varphi_{n}^{\alpha_n}\rangle$ ($\alpha_n=\pm$) are the eigenvalues and eigenstates, respectively of the $n$th Floquet operator $\mathcal{\hat{F}}_n$.

The general form of the eigenstates is quite involved (not shown). However, in the resonant limit  $\tan(\alpha)=-\omega_{R}/\delta\gg1$, we can simplify the Floquet map $\mathcal{\hat{F}}_n$.
The spectral decomposition of  $\mathcal{\hat{F}}_n$ is given as
 \begin{eqnarray}
\label{spectral decomposition}
&&\mathcal{\hat{F}}_n=\exp\lbrace-i\varphi_{n}\rbrace\vert\varphi_{n}^+\rangle\langle\varphi_{n}^+\vert+\exp\lbrace i\varphi_{n}\rbrace\vert\varphi_{n}^-\rangle\langle\varphi_{n}^-\vert.
\end{eqnarray}
Here the quasienergy $\varphi_{n}$ is given by
 $\varphi_{n}=\frac{(\sqrt{\chi_{n}^2+\omega_{0}^2})T}{2}$,
where we introduced the notation $\chi_n=g\sqrt{2I_n/m\omega_r}\cos\theta_n$. The normalized eigenstates are
 $\vert\varphi_{n}^+\rangle=\eta_n\vert0\rangle+\xi_n \vert1\rangle,$ and 
 $\vert\varphi_{n}^-\rangle=\xi_n\vert0\rangle-\eta_n \vert1\rangle$,
where
 $\eta_n=\frac{k_{n}}{\sqrt{1+k_n^2}}$,
 $\xi_n=\frac{1}{\sqrt{1+k_n^2}}$, and $k_n=\frac{\omega_{0}+\sqrt{\chi_{n}^2+\omega_{0}^2}}{\chi_{n}}$.
  \begin{figure}[H]
 \includegraphics[width=0.45\linewidth,height=2.5in]{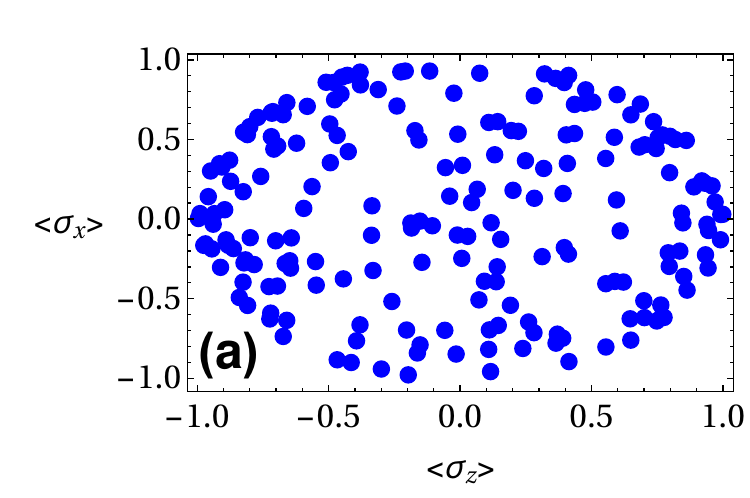}\ \includegraphics[width=0.45\linewidth,height=2.5in]{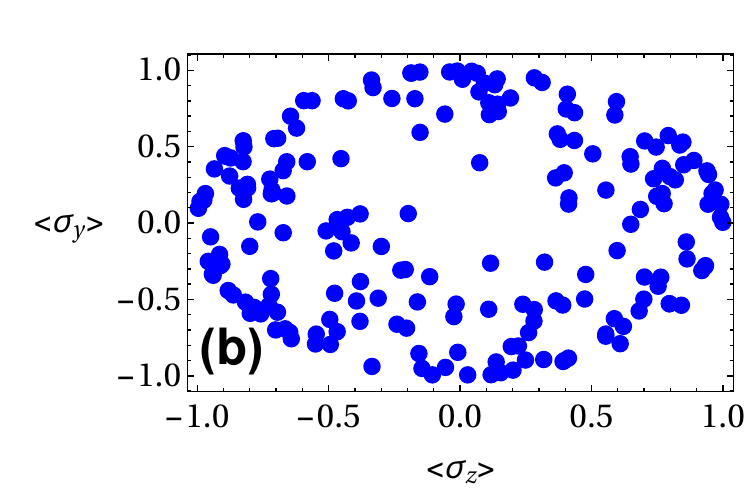}\\
 \includegraphics[width=0.45\linewidth,height=2.8in]{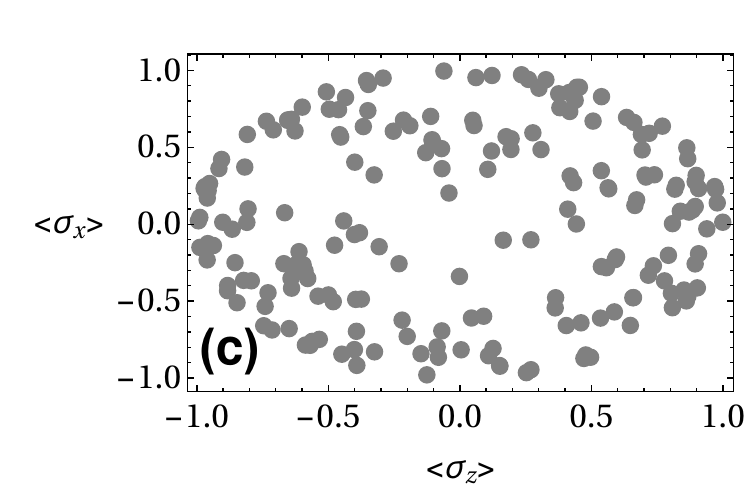}\ \includegraphics[width=0.45\linewidth,height=2.8in]{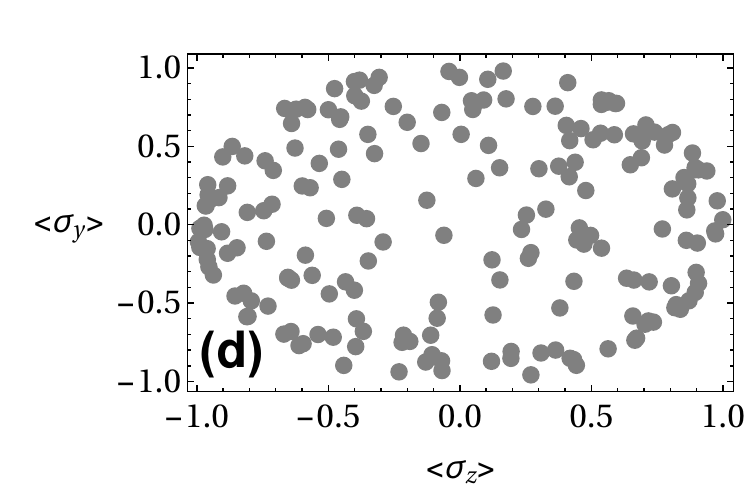}
\caption{Poincar\'{e} sections for  $(\langle\sigma_x\rangle,~\langle\sigma_z\rangle)$ and $(\langle\sigma_y\rangle,~\langle\sigma_z\rangle)$  in the regular regime at $K=0.5$ ( (a) and (b)) and chaotic regime at $K=10$ ((c) and (d)) . The parameters are $m=1$, $g=1$,  $\omega_{0}=1$, $\omega_{r}=0.2$, $T=1$, $\alpha=\pi/2$. The values of the parameters in the real units are $K=\epsilon I_{0} T \frac{6\pi \mu}{m^2\omega_{r}^{2}}$, $\mu=\frac{\omega_{r}^{2}m}{2a_{0}^2}$,
$I_{0}=\frac{m}{2}x_{0}^2 \omega_{r}$, $m=6\times10^{-17}$Kg, $x_{0}=a_{0}=5\times10^{-3}$m, $T=10\mu$s, $\omega_{r}=\omega_{0}=2\pi\times5\times10^{6}$Hz, for chaotic case $\varepsilon=0.003$ and for the regular case $\varepsilon=0.0003$.}
\label{regularx}
 \end{figure}
  Now let us consider that initially the system is prepared in the state $\vert\psi(0)\rangle=\vert0\rangle$.
Therefore, the explicit form of the evolved wave function is calculated as
\begin{eqnarray}\label{time dependent}
&&\vert\psi(t=NT)\rangle=\mathcal{A}_{11}\lbrace\eta_1\exp(-i\varphi_1)(\eta_N\vert0\rangle+\xi_N\vert1\rangle)\rbrace+\mathcal{A}_{12}\lbrace\eta_1\exp(-i\varphi_1)(\xi_N\vert0\rangle\nonumber\\&&-\eta_N\vert1\rangle)\rbrace
+\mathcal{A}_{21}\lbrace \xi_1\exp(i\varphi_1)(\eta_N\vert0\rangle+\xi_N\vert1\rangle)\rbrace+
\mathcal{A}_{22} \lbrace\xi_1\exp(i\varphi_1)(\xi_N\vert0\rangle-\eta_N\vert1\rangle)\rbrace,\nonumber\\
&&\mathcal{A}=\prod\limits_{n=2}^{N}G_{n}\lbrace\varphi\rbrace,~~~\nonumber\\&&
G_{n}\lbrace\varphi\rbrace=\begin{bmatrix}\label{time dependentx}
    \exp(-i\varphi_{n})(\eta_{n}\eta_{n-1}+\xi_{n}\xi_{n-1}) & \exp(-i\varphi_{n})(\eta_{n}\xi_{n-1}-\xi_{n}\eta_{n-1}) \\
    \exp(i\varphi_{n})(\xi_{n}\eta_{n-1}-\eta_{n}\xi_{n-1}) & \exp(i\varphi_{n})(\eta_{n}\eta_{n-1}+\xi_n\xi_{n-1})
\end{bmatrix}.\nonumber\\
\end{eqnarray}
For more details of the analytical solution and normalization of the wave function, we refer to  Appendix~.\ref{appendix1} and Appendix~.\ref{Normalizationx}. Taking into account  Eq.~(\ref{wave function})-Eq.~(\ref{time dependent}) we calculate the expectation values of the spin components $\langle\sigma_\alpha\rangle$, $\alpha=x,y,z$. The explicit formulas are given in the  Appendix~.\ref{spindynamics}.

We note that $\varphi_{n}=\frac{(\sqrt{\chi_{n}^2+\omega_{0}^2})T}{2}$, where $\chi_n=g\sqrt{2I_n/m\omega_r}\cos\theta_n$ and $\left(I_n,~\theta_n\right)$ is described by the map Eq.~(\ref{recurrence relation}).
Therefore, depending on the parameter of stochasticity $K$, the phase $\varphi_{n}$ can be either
non-commensurate and random or smooth and regular.
In the spirit of the work \cite{khomitsky2013spin} we explore the interplay between the chaotic classical (cantilever) and quantum (NV spin) dynamics in the next section. 
 
 \subsection {Expectation values of the NV spin components}
 \label{result}
 We see from Fig.~(\ref{regular}) that the Poincar\'{e} sections for $(I_{n},~\theta_{n})$ clearly distinguish the motion of cantilever in the regular and chaotic regime. Now if we attach a NV center spin to the cantilever, we need to check whether the Poincare sections of the spin dynamics show a contrast in the regular and chaotic regimes of the cantilever or not. For this purpose we plot the Poincar\'{e} section of $(\langle\sigma_x\rangle,~\langle\sigma_z\rangle)$ and $(\langle\sigma_y\rangle,~\langle\sigma_z\rangle)$ in Fig.~\ref{regularx}(a) and (b) when cantilever performs motion in regular regime and (c) and (d) when cantilever performs motion in chaotic regime.  We fail to distinguish the effects due to regular and chaotic regions in the Poincar\'{e} sections of the spin dynamics of the NV center.

\begin{figure}[H]
\centering
 \includegraphics[width=0.45\linewidth,height=2in]{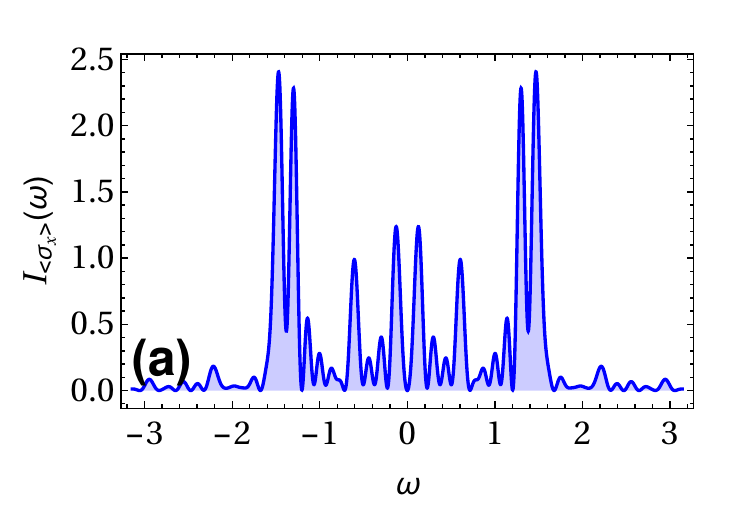}\ \includegraphics[width=0.45\linewidth,height=2in]{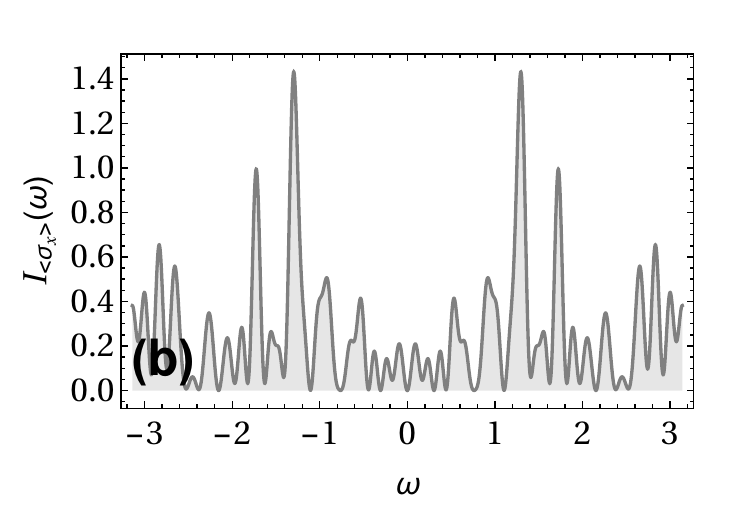}\\
 \includegraphics[width=0.45\linewidth,height=2in]{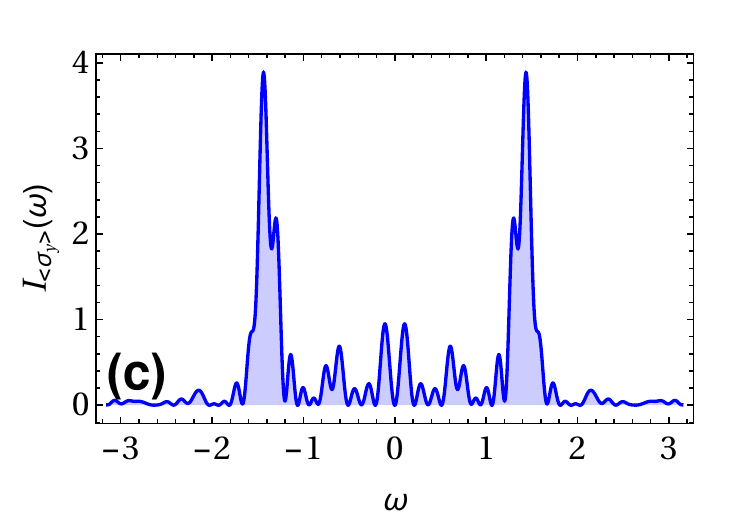}\ \includegraphics[width=0.45\linewidth,height=2 in]{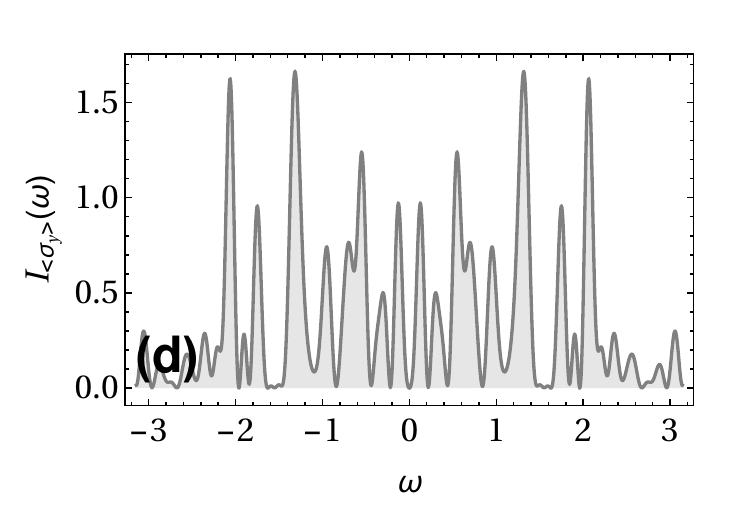}\\
 \includegraphics[width=0.45\linewidth,height=2in]{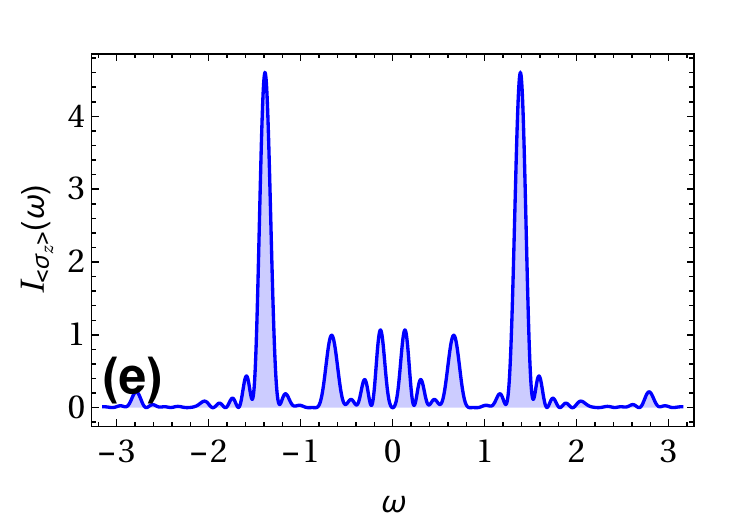}\ \includegraphics[width=0.45\linewidth,height=2in]{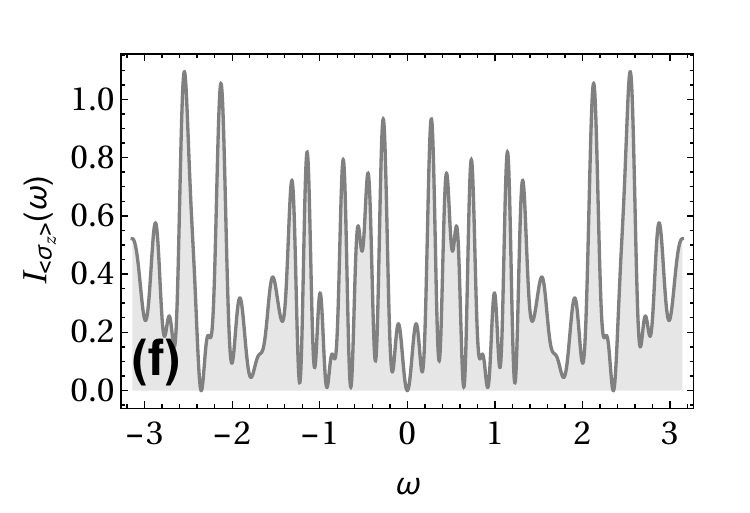}
\caption{Fourier Power spectrum density for expectation values of $\sigma_{x,y,z}$ in the regular regime ((a), (c) and (e)) at
$K=0.5$ (Blue), and in the chaotic regime ((b), (d) and (f)) at $K=10$ (Gray). The parameters used for the plot are $m=1$, $g=1$,  $\omega_{0}=1$, $\omega_{r}=0.2$, $T=1$, $\alpha=\pi/2$. The values of the parameters in the real units: $K=\epsilon I_{0} T \frac{6\pi \mu}{m^2\omega_{r}^{2}}$, $\mu=\frac{\omega_{r}^{2}m}{2a_{0}^2}$ $I_{0}=\frac{m}{2}x_{0}^2 \omega_{r}$, $m=6\times10^{-17}$Kg, $x_{0}=a_{0}=5\times10^{-3}$m, $T=10\mu$s, $\omega_{r}=\omega_{0}=2\pi\times5\times10^{6}$Hz, for chaotic case $\epsilon=0.003$ and for the regular case $\epsilon=0.0003$.}
\label{psdsigmax}
 \end{figure}

\begin{figure}[H]
 \includegraphics[width=0.45\linewidth,height=2.2in]{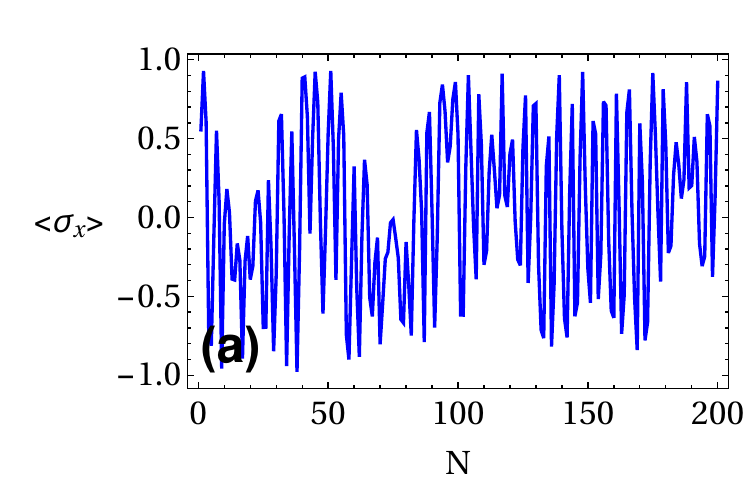}\ \includegraphics[width=0.45\linewidth,height=2.2in]{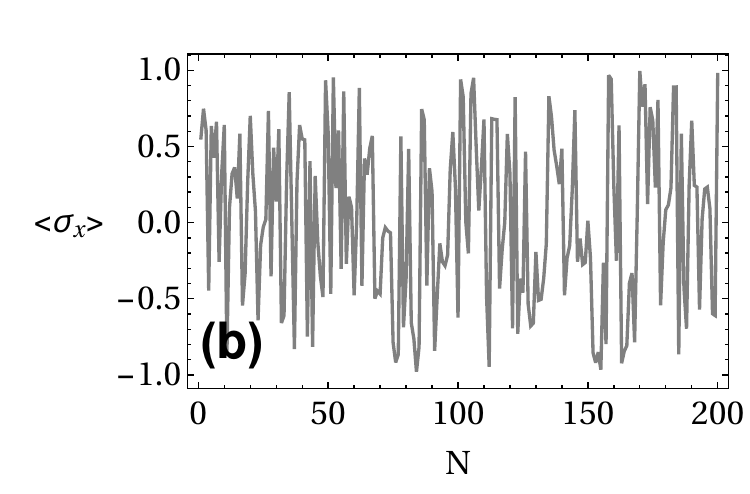}\\
 \includegraphics[width=0.45\linewidth,height=2.2in]{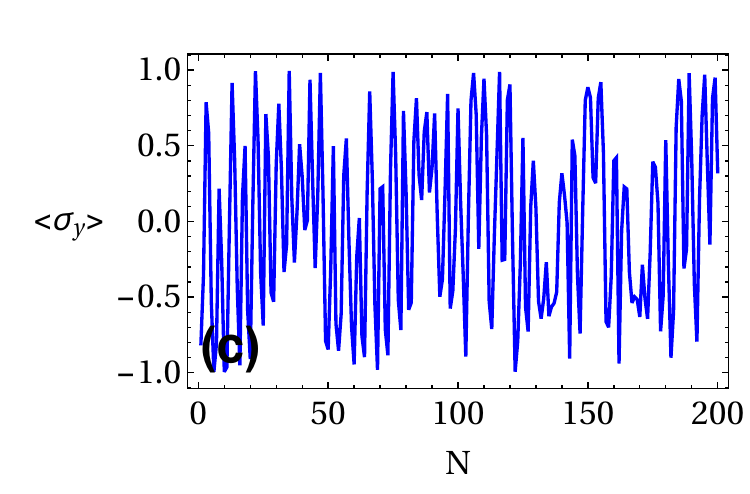}\ \includegraphics[width=0.45\linewidth,height=2.2in]{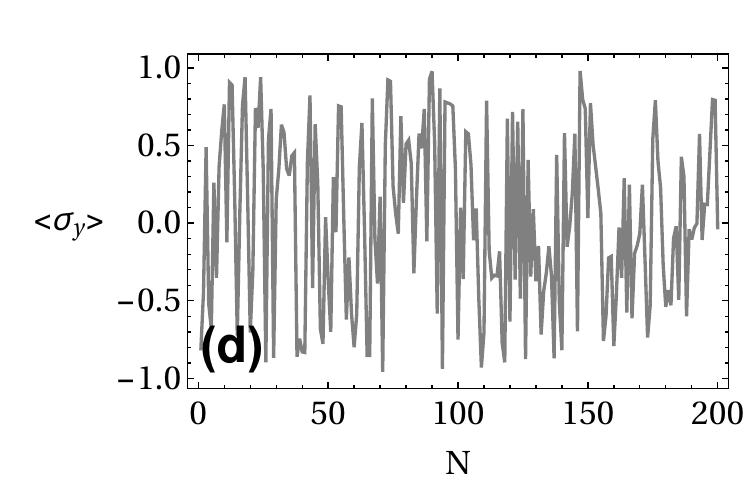}\\
  \includegraphics[width=0.45\linewidth,height=2.2in]{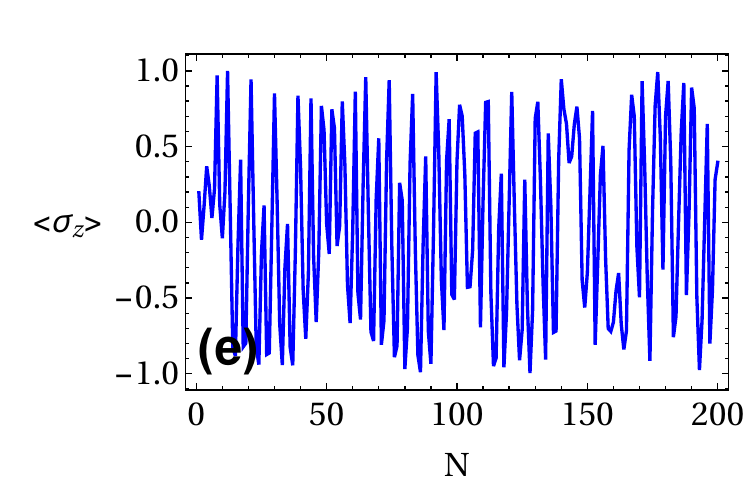}\ \includegraphics[width=0.45\linewidth,height=2.2in]{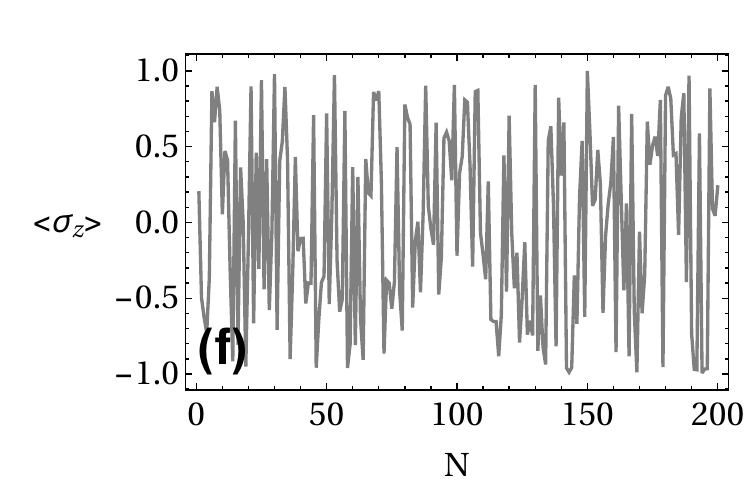}
\caption{Spin dynamics  for  $\langle\sigma_x\rangle$, $~\langle\sigma_y\rangle$ and $\langle\sigma_z\rangle$  in the regular regime at $K=0.5$ (see (a), (c) and (e)) and $\langle\sigma_x\rangle$, $~\langle\sigma_y\rangle$ and $\langle\sigma_z\rangle$ in the chaotic regime at $K=10$ (see (b), (d) and (f)). The parameters are $m=1$, $g=1$,  $\omega_{0}=1$, $\omega_{r}=0.2$, $T=1$, $\alpha=\pi/2$. The values of the parameters in the real units: $K=\epsilon I_{0} T \frac{6\pi \mu}{m^2\omega_{r}^{2}}$, $\mu=\frac{\omega_{r}^{2}m}{2a_{0}^2}$, $I_{0}=\frac{m}{2}x_{0}^2 \omega_{r}$, $m=6\times10^{-17}$Kg, $x_{0}=a_{0}=5\times10^{-3}$m, $T=10\mu$s, $\omega_{r}=\omega_{0}=2\pi\times5\times10^{6}$Hz, for chaotic $\varepsilon=0.003$ and for regular $\varepsilon=0.0003$.}
\label{regularyx}
 \end{figure}

 The Poincar\'{e} sections of the spin dynamics evolve more or less in the same manner for both the regular and the chaotic cases (see Fig.~\ref{regularx}).
 In order to delve deeper to identify the differences, we calculate the Fourier power spectrum for observances defined as $I_{x,y,z}=\bigg\vert\int\limits_{-\infty}^{\infty}\langle\sigma_{x,y,z}\rangle\exp(-i\omega t)dt\bigg\vert^{2}$. The Fourier power spectrum 
as shown in Fig.~\ref{psdsigmax} displays differences in the regular and chaotic regime. We see that when stochasticity parameter varies from  $K=0.5$ (regular) to $K=10$ (chaotic), the broadness of power spectrum increases. It is much broader in the chaotic case as compared to the regular case. The broadening of spectrum is a signature of chaos
which sets in our system for $K>1$. We see that the Fourier spectra of all spin components $\langle\sigma_{x,y,z}\rangle$, are broadened. To see the behavior of spin dynamics, we plot the time dependence of different spin components. While $\langle\sigma_{x,y,z}\rangle$ components perform fast, chaotic oscillations in chaotic regime (see Figs.~\ref{regularyx} (b), (d) and (f)), they show quasi-periodic oscillation in the regular regime (see Figs.~\ref{regularyx} (a), (c) and (e)).
 \begin{figure}[H]
 \includegraphics[width=0.45\linewidth,height=2.2in]{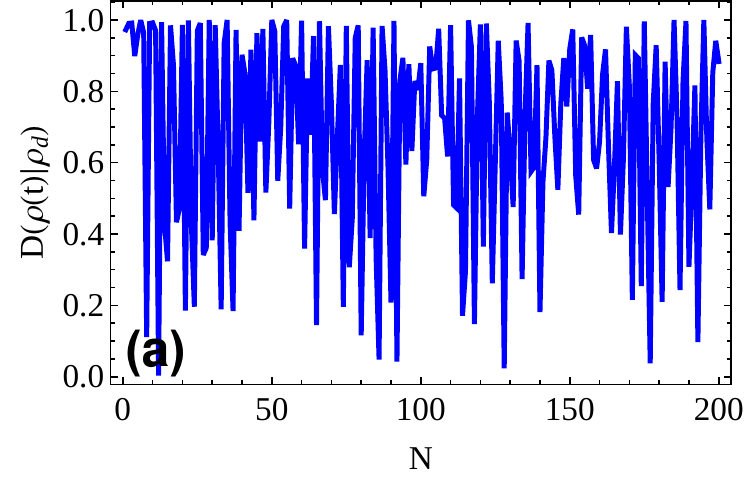}\ \includegraphics[width=0.45\linewidth,height=2.2in]{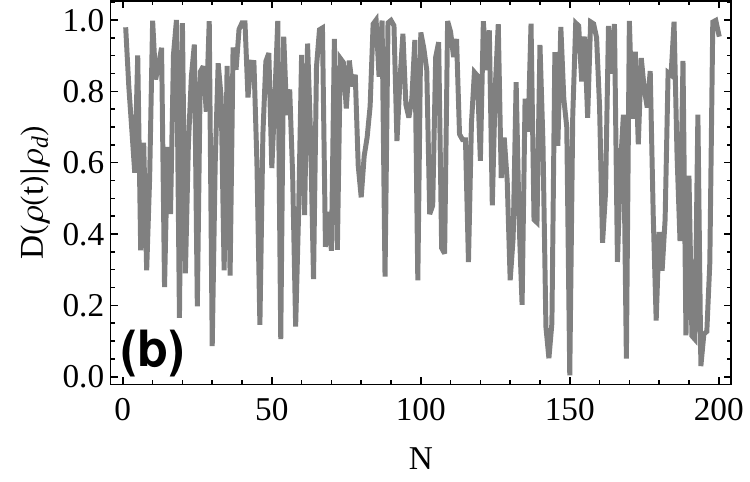}
\caption{The quantum coherence in (a) the regular regime at $K=0.5$ (Blue) and (b) the chaotic regime at $K=10$ (Gray), for
the following values of the parameters $m=1, g=1$,  $\omega_{0}=1$, $\omega_{r}=0.2$, $T=1$, $\alpha=\pi/2$. The values of parameters are $K=\epsilon I_{0} T \frac{6\pi \mu}{m^2\omega_{r}^{2}}$, $\mu=\frac{\omega_{r}^{2}m}{2a_{0}^2}$ $I_{0}=\frac{m}{2}x_{0}^2 \omega_{r}$, $m=6\times10^{-17}$Kg, $x_{0}=a_{0}=5\times10^{-3}$m, T=10$\mu$s, $\omega_{r}=\omega_{0}=2\pi\times5\times10^{6}$Hz, for chaotic $\epsilon=0.003$ and for regular $\epsilon=0.0003$.  }
\label{chaoticx}
 \end{figure}
 
 \subsection{Quantum coherence}
Quantum coherence is the resource for performing vast number of quantum information protocols. In many-body system  the quantum coherence is the essence of entanglement and plays an important role in understanding some physical phenomena of quantum information and quantum optics. Relative entropy and l1-norm measures are also a monotone of coherence \cite{RevModPhys.89.041003}. By incoherent operations one can generate  coherence that quantifies  maximal entanglement \cite{PhysRevLett.115.020403}. The loss of coherence in a quantum system may happen due to two different reasons: In one case when the system is in contact with the environment or a thermal bath, the coupling to the environment may cause decoherence, which is a stochastic phenomenon.  In the other case, the coupling of a quantum system with a classical chaotic system, may lead to a loss of coherence. We focus on the second case where dynamical chaos due to the non-linearity in the classical system \cite{zou2012phase} may result a loss of coherence.  Here we explore the problem of generation of coherence for the NV spin coupled to a nanocantilever in a regular or a chaotic regime.

In particular, we prepare the NV center initially in a mixed state:
\begin{eqnarray}\label{mixed state}
\hat{\rho}(0) &=& p_{1}\vert 0 \rangle\langle0|+p_{2}\vert 1 \rangle\langle1|.
\end{eqnarray}
The time evolved density matrix is given by evolution operator Eq.~(\ref{Schrodinger}) as:
\begin{eqnarray}
\hat{\rho}(t)=(\mathcal{\hat{U}}^{N})^{-1}\hat{\rho}(0)\mathcal{\hat{U}}^{N}, 
\end{eqnarray}
\begin{eqnarray}\label{densityformalism}
&&\hat{\rho}(t)=\rho_{11}\vert0\rangle\langle0\vert+\rho_{12}\vert0\rangle\langle1\vert+\rho_{21}\vert1\rangle\langle0\vert+\rho_{22}\vert1\rangle\langle1\vert .
\end{eqnarray}
The elements of the time-evolved density matrix are given in the  Appendix~.\ref{Quantumcoherence}, where all the elements of $\hat{\rho}(t)$ are time-dependent.
We quantify the quantum coherence in terms of the relative entropy as
\begin{eqnarray}\label{divergence}
&&\mathcal{D}\big(\hat{\rho}(t)|\hat{\rho}_d(t)\big)=Tr\{\hat{\rho}(t)\ln{\hat{\rho}(t)}-\hat{\rho}(t)\ln \hat{\rho}_{d}(t)\}.
\end{eqnarray}
Here $\hat{\rho}_d(t)$ is the diagonal part of $\hat{\rho}(t)$.
The eigenvalues of the density matrix $\rho(t)$ are:
$E_{\pm}=\frac{1}{2}\Big(\rho_{11}+\rho_{22}\pm\sqrt{\rho_{11}^2+\rho_{22}^2+4 \rho_{21}\rho_{12}-2\rho_{11}\rho_{22}}\Big)$.
Now, taking into account Eq.~(\ref{time dependent}), we calculate quantum coherence in terms of relative entropy as:
\begin{eqnarray}\label{analytic result for quantum coherence}
&&\mathcal{D}\big(\hat{\rho}(t)|\hat{\rho}_d(t)\big)=E_{+}\ln{E_{+}}+E_{-}\ln{E_{-}}-\rho_{11}\ln{\rho_{11}}-\rho_{22}\ln{\rho_{22}}.
\end{eqnarray}
The stochasticity parameter $K$ appears in the expression of $\rho(t)$ as $\eta_{N}$, $\xi_{N}$ which contain $I_{n}$ and $\theta_{n}$. The relative entropy $\mathcal{D}$
for regular and chaotic cases are plotted in Fig.~\ref{chaoticx}. We see that quantum coherence in regular case is doing quasi-periodic oscillation  while in chaotic regime coherence varies abruptly. This observation supports the fact that the chaos destroys the quantum coherence.

\subsection{Quantum Poincar\'{e} recurrence}
\textit{"Any phase-space configuration $(I,\theta)$ of a system enclosed in a finite volume will be repeated as accurately as one wishes after a finite interval of time"}. This statement is the essence of the Poincar\'{e} recurrence theorem and holds in the quantum case  also \cite{bocchieri1957quantum}. Any time-dependent periodic Hamiltonian would reunite itself infinitely often over time. Suppose the system has a continuous energy spectrum corresponding to the classical systems, then the quantum recurrence theorem does not hold. A quantum system that is bounded defined by a Hamiltonian $H_{0}$ has a discrete spectrum when subjected to a nonresonant time-dependent periodic potential $V$ with $V(t) =V(t +\tau)$ for an arbitrary period $\tau$. For any initial configuration of the system, both the wave function and the energy reunite itself over time \cite{PhysRevLett.48.711}. The time passed off during the recurrence is known as Poincar\'{e} recurrence time. The Quantum Poincar\'{e} recurrence means that the distance between the initial and evolved states can become smaller than the characteristic $\epsilon$:
$\Vert\phi(t)-\phi(0)\Vert<\epsilon$.
\begin{figure}[t!]
 \includegraphics[width=0.45\linewidth,height=2.4in]{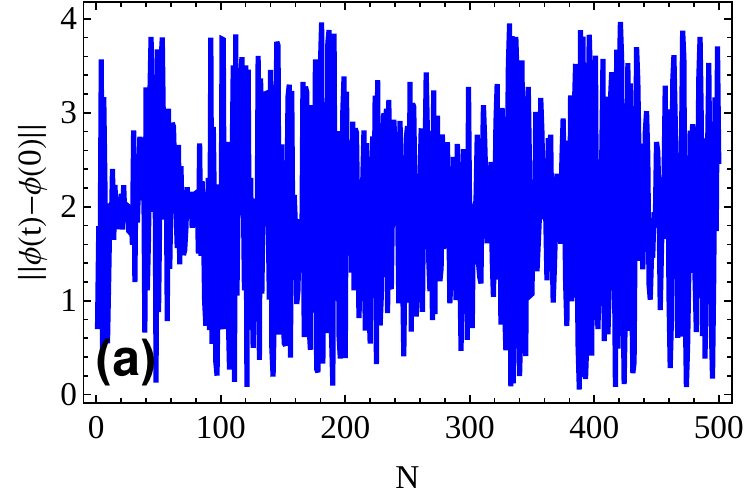}\ \includegraphics[width=0.45\linewidth,height=2.4in]{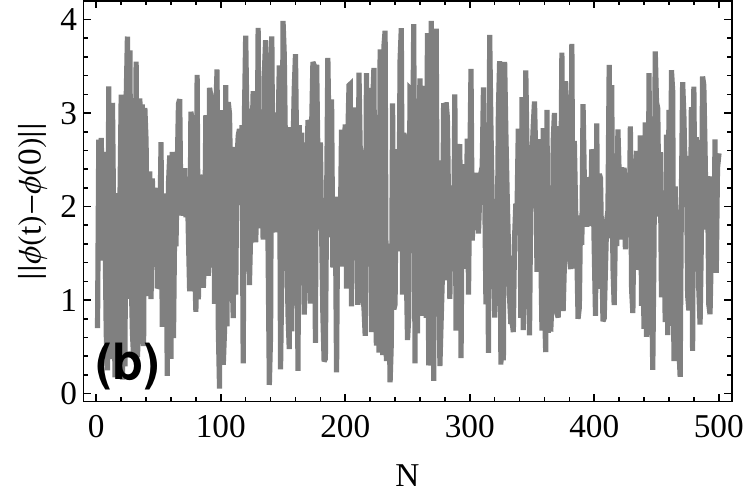}
\caption{Quantum Poincar\'{e} recurrence as a function of time
(i.e. number of kicks) in (a) the regular regime at $K=0.5$ (Blue) and (b) the chaotic regime at $K=10$ (Gray). The parameters are $m=1$, $g=1$, $\omega_{0}=1$,  $\omega_{r}=0.2$, $T=1$, $\alpha=\pi/2$. The values of parameters: $K=\epsilon I_{0} T \frac{6\pi \mu}{m^2\omega_{r}^{2}}$, $\mu=\frac{\omega_{r}^{2}m}{2a_{0}^2}$ $I_{0}=\frac{m}{2}x_{0}^2 \omega_{r}$, $m=6\times10^{-17}$Kg, $x_{0}=a_{0}=5\times10^{-3}$m, T=10$\mu$s, $\omega_{r}=\omega_{0}=2\pi\times5\times10^{6}$Hz, for chaotic $\epsilon=0.003$ and for regular $\epsilon=0.0003$ . }
\label{QuantumPoincareRecurrence}
 \end{figure}
Taking  Eq.~(\ref{time dependent}) into account the explicit expression for the distance of the time evolved state from the initial state is:
\begin{eqnarray}
&&\Vert\phi(t)-\phi(0)\Vert=2-\Big(A_{11}^{*}{\eta_{1}^{*}\exp(i \varphi_{1})\eta_{N}^{*}}+A_{12}^{*}{\eta_{1}^{*}\exp(i \varphi_{1})\xi_{N}^{*}}+\nonumber\\&&A_{21}^{*}\xi_{1}^{*}\exp(-i \varphi_{1})\eta_{N}^{*}+A_{22}^{*}\xi_{1}^{*}\exp(-i
\varphi_{1})\xi_{N}^{*}+A_{11}\eta_{1}\exp(-i \varphi_{1})\eta_{N}\nonumber\\&&+A_{12}\eta_{1}\exp(-i \varphi_{1})\xi_{N}+A_{21}\xi_{1}\exp(i \varphi_{1})\eta_{N}+A_{22}\exp(i \varphi_{1})\xi_{N}\Big).
\end{eqnarray}

The above expression of quantum Poincar\'{e} recurrences is plotted for the regular $K<1$ and chaotic cases $K>1$ separately in Fig. \ref{QuantumPoincareRecurrence} (a) and (b), respectively.  From these figures we see a slight difference in behavior of the system in two regimes. In the regular case Fig.~\ref{QuantumPoincareRecurrence} (a) we see a  trend of quasiperiodic modulation of the amplitude, while in  the chaotic case Fig.~\ref{QuantumPoincareRecurrence} (b), the distance measure between the wave functions is the essence of a noise.
Analyses of the recurrence show the absence of the exponential decay of Poincar\'{e} recurrence, while the exponential decay is a hallmark of quantum chaos
\cite{casati1999quantum}. 
The effect we observe in our system is non-conventional for quantum chaos. The reason for the absence of the conventional quantum chaos phenomenon is the low dimensionality of the spin space. On the other hand, chaotic dynamics of cantilever plays the role of external noise for NV spin and has a stochastic character rather than a dynamical. It destroys the nature of quasiperiodic revivals in spin dynamics, and quantum recurrence becomes a random event. The dynamics of the quantum system is distinct from the behavior of the regular systems. 
Therefore, we term this effect as hybrid quantum-classical chaos.

One of the interesting characteristics of the hybrid quantum-classical chaos is the time-translation symmetry breaking (TTSB). The Hamiltonian $\hat{H}_n$ Eq.~(\ref{effective Hamiltonian}), taken at different times form a set of noncommuting Hamiltonians: ${\hat{H}_n }$. The integer $n$ defines discrete moment of time $t_n=nT$, where $T$ is the period between the pulses applied to the cantilever. Therefore, ${\hat{H}_n}$ is a set of elements repeated in time $\hat{H}_n(I_n\,\theta_n) \equiv \hat{H}_{n+k}(I_{n+k},\,\theta_{n+k})$, when canonical variables repeat their values
$(I_{n+k},\,\theta_{n+k})=(I_n,\,\theta_n)$ i.e., the Floquet time crystal
\cite{PhysRevLett.109.160401,PhysRevLett.117.090402}. On the other hand the quantum Poincar\'{e} recurrence occurs if the distance between state vectors is small
$\Vert\phi((n+k)T)-\phi(nT)\Vert<\epsilon$, where $\epsilon$ is the characteristic small parameter of the recurrence. The time-translation symmetry underlies conservation of energy, reproducibility of the wave function and Hamiltonian. TTSB occurs if for each $t_n$ and for every state $\vert\phi(nT)\rangle$  there exists an operator $\mathcal{A}$ for which  at least one of the two conditions $\hat{H}_n(I_n,\,\theta_n) \equiv \hat{H}_{n+k}(I_{n+k},\,\theta_{n+k})$ and $\langle\vert\phi(nT)\vert\mathcal{A}\vert\vert\phi(nT)\rangle=\langle\vert\phi((n+k)T)\vert\mathcal{A}\vert\vert\phi((n+k)T)\rangle$ is violated. In our case operator $\mathcal{A}$ corresponds to the spin operator $\mathcal{A}\equiv\hat S$. We note that the conditions $\hat{H}_n(I_n,\,\theta_n) \equiv \hat{H}_{n+k}(I_{n+k},\,\theta_{n+k})$, and $(I_{n+k},\,\theta_{n+k})=(I_n,\,\theta_n)$ hold only in the regular case (elliptic trajectories) and are violated in the chaotic case when invariant torus are destroyed and dynamics is not periodic in the phase space. TTSB occurs due to the hybrid character of quantum classical chaos, meaning that Quantum Poincar\'{e} recurrence of the wave function holds while the periodicity of the Hamiltonian not.

\begin{figure}[H]
 \includegraphics[width=0.8\linewidth,height=3in]{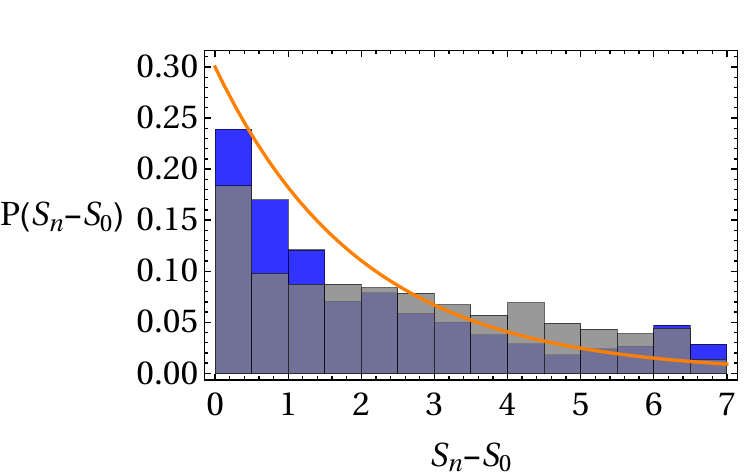} 
\caption{Histogram plot of level statistics of Hamiltonian $\hat{H}_n$ in the regular regime at $K=0.5$ (Blue) and in the chaotic regime $K=10$ (Gray). A reference plot for poissonian statistics (Orange) is also shown. For this plot we have taken upto 1000 kicks to get the ensemble.
The parameters are $m=1$, $g=1$, $\omega_{0}=1$,  $\omega_{r}=0.2$, T=1, $\alpha=\pi/2$. The values of parameters: $K=\epsilon I_{0} T \frac{6\pi \mu}{m^2\omega_{0}^{2}}$, $\mu=\frac{\omega_{r}^{2}m}{2a_{0}^2}$ $I_{0}=\frac{m}{2}x_{0}^2 \omega_{r}$, $m=6\times10^{-17}$Kg, $x_{0}=a_{0}=5\times10^{-3}$m, T=10$\mu$s, $\omega_{r}=\omega_{0}=2\pi\times5\times10^{6}$Hz, for chaotic case $\epsilon=0.003$ and for the regular case $\epsilon=0.0003$. }
\label{Hamiltonianlevestatistics}
 \end{figure}

\subsection{Level statistics for spin-1/2 case}\label{Level_statistics_1/2}

The eigenvalues of the the  Hamiltonian $\hat{H}_n$ ( Eq.~(\ref{effective Hamiltonian})) are given by:
$E^{(n)}_{1,2}=\pm\frac{1}{2}\sqrt{\chi_{n}^2+\omega_{0}^2}$,
where  $\chi_{n}=g\sqrt{\frac{2I_{n}}{m \omega_{0}}}\cos{\theta_{n}}$.
Each Hamiltonian from the set $\{\,\hat{H}_n\,\}$ has two energy
levels. We explore the distances between the levels:
\begin{eqnarray}\label{distance between levels}
&& S_n=E_1^{(n)}-E_2^{(n)}=\sqrt{\omega_{0}^2+g^2\Big(\frac{2I_{n}}{m\omega_{0}}\Big)\cos^2{\theta_{n}}},
\end{eqnarray}
for each Hamiltonian and construct the distribution functions $P(S_n-S_{0})$ for regular $K<1$ and chaotic $K>1$ cases. Here $S_{n}$ is the separation between two energy levels
and $S_0$ corresponds to the maximum of $P(S_0)$. We see that the level statistics is Poissonian in the both regular $K<1$ and chaotic $K<1$ cases. 
Comparing results of spin dynamics Figs.~\ref{psdsigmax} and \ref{regularyx}, with level statistics Fig.~ \ref{Hamiltonianlevestatistics} we see that in the both cases level statistics is of Poissonian type, while we expect it to be Gaussian in chaotic case \cite{khomitsky2016regular}. Thus for spin 1/2 case, in spite of the chaotic quantum spin dynamics we do not observe statistical characteristics of quantum chaos.

\section{Dynamics of a three-level NV system}\label{SPIN1}
We proceed to analyse a more general case and consider a three-level NV center.
The effective Hamiltonian of the NV center for spin $S=1$ attached to the cantilever can be written as: 
\begin{eqnarray}\label{effective Hamiltonian1}
\hat{H}_{n}=\hat{H}_{NV}+g\hat{V}_{c,NV},
\end{eqnarray}
where $\hat H_{NV}=\sum\limits_{i=\pm 1}\left(-\delta_i\vert i\rangle\langle i\vert+\frac{\Omega_i}{2}(\vert 0\rangle\langle i\vert+\vert i\rangle\langle 0\vert)\right)$ is the Hamiltonian of the NV center  \cite{Rabl2009} and $\hat{V}_{c,NV}=\sqrt{2I_n/m\omega_r}\cos\theta_n\hat{S}_{z,NV}$ is the coupling term with the nonlinear cantilever. For spin $S=1$, $S_{z,NV}=\frac{1}{2}(\cos{\alpha}S_{z}+\sin{\alpha}S_{x})$, where $S_{x,z}=\sigma_{x,z}(1)$ is spin components for $S=1$ case. For numerical calculations we consider $\delta_{\pm1}=\delta=1$ and $\Omega_{\pm}=\Omega=1$. Similar to the analysis of spin-1/2 case discussed in Section~ \ref{spin_dynamics}, we calculate time-dependent wave function for the system Eq.~(\ref{effective Hamiltonian1}) using $\delta_{\pm}=\delta$ and $\Omega_{\pm}=\Omega$ can be written as:
 \begin{figure}[t!]
\includegraphics[width=0.45\linewidth,height=2.2in]{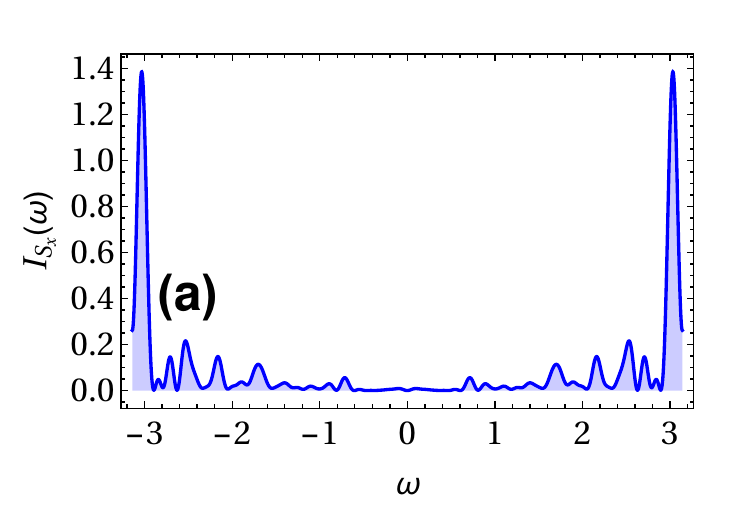}\ \includegraphics[width=0.45\linewidth,height=2.2in]{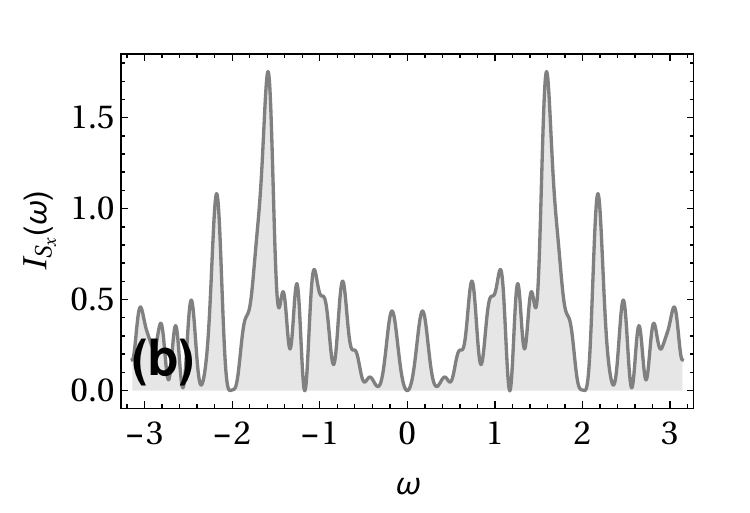}\\
 \includegraphics[width=0.45\linewidth,height=2.2in]{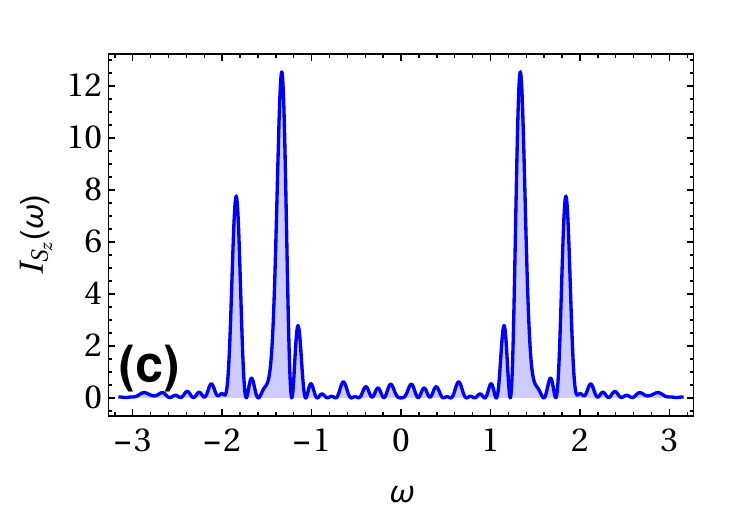}\ \includegraphics[width=0.45\linewidth,height=2.2in]{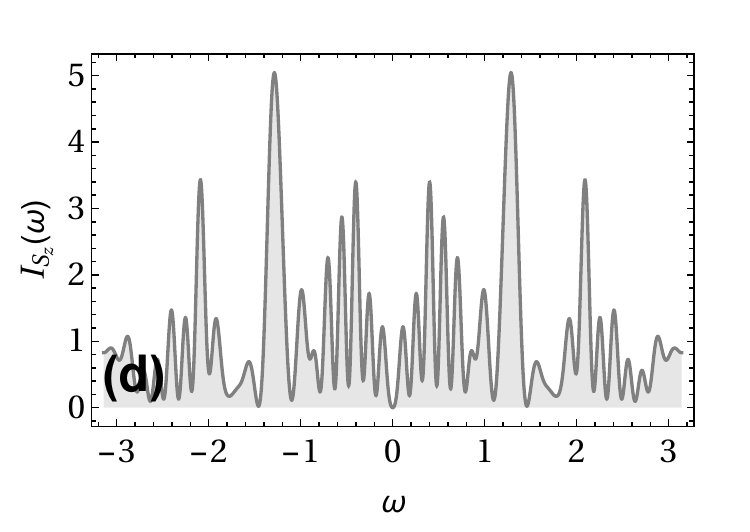}
\caption{Fourier Power spectrum density  for the components $S_{x,z}$ in the regular regime ((a) and (c)) at
$K=0.5$ (Blue), and in the chaotic regime ((b) and (d)) at $K=10$ (Gray). The parameters used for the plot are $m=1$, $g=1$,  $\Omega=1$, $\delta=1$, $\omega_{r}=0.2$, $T=1$, $\alpha=\pi/2$. The values of the parameters in the real units: $K=\epsilon I_{0} T \frac{6\pi \mu}{m^2\omega_{r}^{2}}$, $\mu=\frac{\omega_{r}^{2}m}{2a_{0}^2}$ $I_{0}=\frac{m}{2}x_{0}^2 \omega_{r}$, $m=6\times10^{-17}$Kg, $x_{0}=a_{0}=5\times10^{-3}$m, $T=10\mu$s, $\omega_{r}=\omega_{0}=2\pi\times5\times10^{6}$Hz, for chaotic case $\epsilon=0.003$ and for the regular case $\epsilon=0.0003$.}
\label{psdsigmaxspin1}
 \end{figure}
\begin{eqnarray}
\label{wave function Spin-1}
\vert\psi(t=NT)\rangle=\sum\limits_{\lbrace\alpha_n\rbrace=1,2,3}\left\lbrace\prod\limits_{n=2}^{N}e^{-i\alpha_n\varphi_n}
 \big\langle \varphi_{n}^{\alpha_n}\big\vert\varphi_{n-1}^{\alpha_{n-1}}\big\rangle
 \right\rbrace e^{-i\alpha_1\varphi_1}\big\langle \varphi_{1}^{\alpha_1}\big\vert\psi(0)\big\rangle\big\vert\varphi_{N}^{\alpha_N}\big\rangle.\nonumber\\
\end{eqnarray}
 Here  ${\alpha_n}\varphi_{n}$ and $\vert\varphi_{n}^{\alpha_n}\rangle$ ($\alpha_n=1,2,3$) are the eigenvalues and eigenstates of the $n$th Floquet operator $\mathcal{\hat{F}}_n$, respectively. 
The spectral decomposition of  $\mathcal{\hat{F}}_n$ is given as
 \begin{eqnarray}
\label{spectral decomposition spin-1}
&&\mathcal{\hat{F}}_n=\exp\lbrace-i\varphi_{n}^{1}\rbrace\vert\varphi_{n}^1\rangle\langle\varphi_{n}^1\vert+\exp\lbrace-i\varphi_{n}^{2}\rbrace\vert\varphi_{n}^2\rangle\langle\varphi_{n}^2\vert+\exp\lbrace - i\varphi_{n}^{3}\rbrace\vert\varphi_{n}^3\rangle\langle\varphi_{n}^3\vert.
\end{eqnarray}
In the above equation, the quasienergy $\varphi_{n}$ is given by
 $\varphi_{n}^{1,2,3}=\delta, \frac{1}{2}(-\delta\pm\sqrt{\delta^{2}+(2\chi_{n}+\sqrt{2}\Omega)^2})T$,
where the notation $\chi_n=g\sqrt{2I_n/m\omega_r}\cos\theta_n$ is already defined in section~\ref{spin_dynamics}. In this case the normalized eigenstates are
 $\vert\varphi_{n}^1\rangle=-\eta_n\vert0\rangle+\xi_n \vert1\rangle+\zeta_n \vert2\rangle,$ and 
 $\vert\varphi_{n}^2\rangle=x_n\vert0\rangle+y_n \vert1\rangle+z_n \vert2\rangle$and
 $\vert\varphi_{n}^3\rangle=u_n\vert0\rangle+v_n \vert1\rangle+w_n \vert2\rangle$. 
The normalization constants of the eigenstates are defined in Appendix~\ref{wavefunctionforspin-1}.

  \begin{figure}[t!]
 \includegraphics[width=0.45\linewidth,height=2.2in]{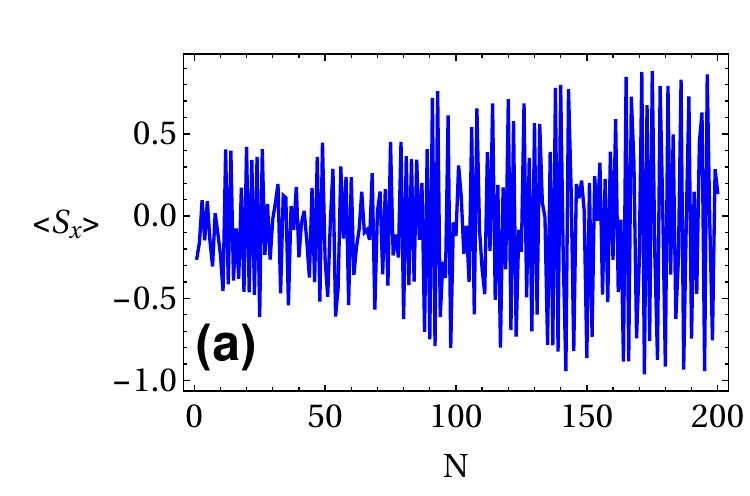}\ \includegraphics[width=0.45\linewidth,height=2.2in]{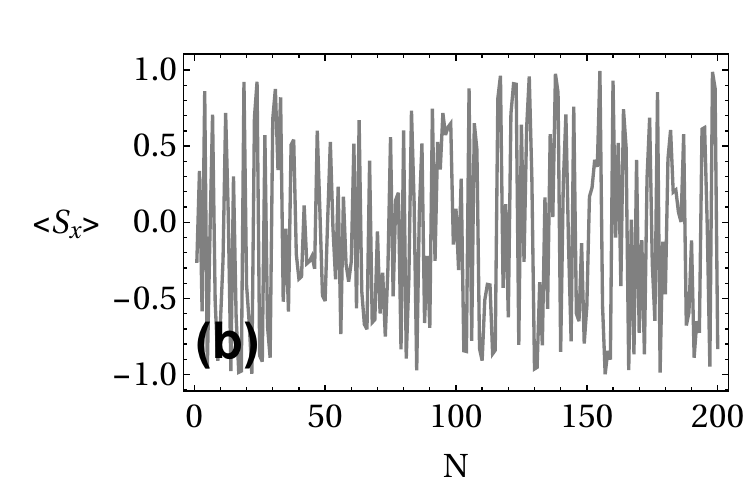}\\
 \includegraphics[width=0.45\linewidth,height=2.2in]{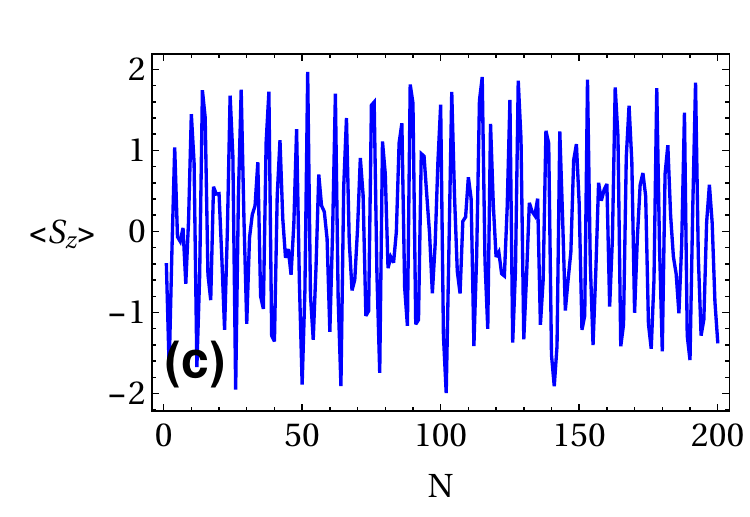}\ \includegraphics[width=0.45\linewidth,height=2.2in]{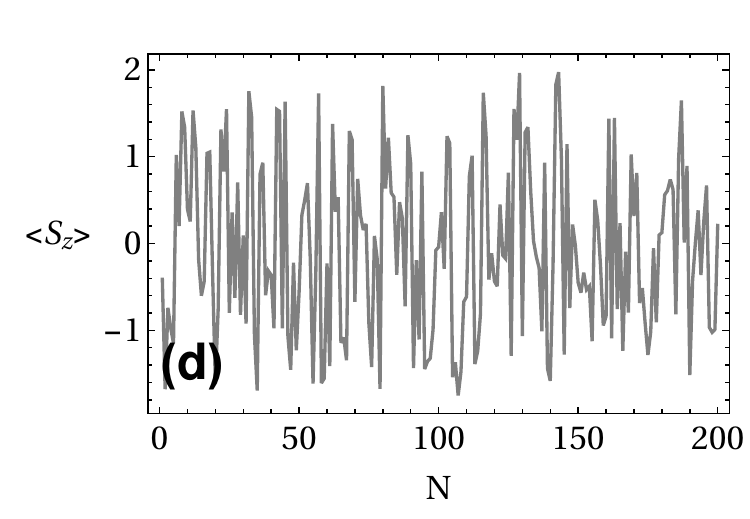}
\caption{Spin dynamics  for the components $S_x$ and $S_z$ for Spin-1 case  in the regular regime at $K=0.5$ ((a) and (c)) and in the chaotic regime at $K=10$ ((b) and (d)). The parameters are $m=1$, $g=1$,   $\Omega=1$, $\delta=1$, $\omega_{r}=0.2$, $T=1$, $\alpha=\pi/2$. The values of the parameters in the real units: $K=\epsilon I_{0} T \frac{6\pi \mu}{m^2\omega_{r}^{2}}$, $\mu=\frac{\omega_{r}^{2}m}{2a_{0}^2}$, $I_{0}=\frac{m}{2}x_{0}^2 \omega_{r}$, $m=6\times10^{-17}$Kg, $x_{0}=a_{0}=5\times10^{-3}$m, $T=10\mu$s, $\omega_{r}=\omega_{0}=2\pi\times5\times10^{6}$Hz, for chaotic $\varepsilon=0.003$ and for regular $\varepsilon=0.0003$.}
\label{regularyxspin1}
 \end{figure}
We prepare the system initially in the state $\vert\psi(0)\rangle=\vert0\rangle$. The explicit form of the evolved wave function from Eq.~(\ref{wave function Spin-1}) takes the form
\begin{eqnarray}\label{time dependent spin-1}
&&\vert\psi(t=NT)\rangle=\mathcal{A}_{11}\lbrace-\eta_1\exp(-i\varphi_1^{1})(-\eta_N\vert0\rangle+\xi_N\vert1\rangle+\eta_N\vert2\rangle)\rbrace\nonumber\\&&+\mathcal{A}_{12}\lbrace-\eta_1\exp(-i\varphi_1^{1})(x_N\vert0\rangle+y_N\vert1\rangle+z_N\vert2\rangle)\rbrace+\mathcal{A}_{13}\lbrace-\eta_1\exp(-i\varphi_1^{1})(u_N\vert0\rangle\nonumber\\&&+v_N\vert1\rangle+w_N\vert2\rangle)\rbrace
+\mathcal{A}_{21}\lbrace x_1\exp(i\varphi_1^{2})(-\eta_N\vert0\rangle+\xi_N\vert1\rangle+\eta_N\vert2\rangle)\rbrace+
\mathcal{A}_{22}\lbrace x_1\exp(i\varphi_1^{2})(x_N\vert0\rangle\nonumber\\&&+y_N\vert1\rangle+z_N\vert2\rangle)\rbrace+
\mathcal{A}_{23}\lbrace x_1\exp(i\varphi_1^{2})(u_N\vert0\rangle+v_N\vert1\rangle+w_N\vert2\rangle)\rbrace+\mathcal{A}_{31}\lbrace u_1\exp(i\varphi_1^{3})(-\eta_N\vert0\rangle\nonumber\\&&+\xi_N\vert1\rangle+\eta_N\vert2\rangle)\rbrace+\mathcal{A}_{32}\lbrace u_1\exp(i\varphi_1^{3})(x_N\vert0\rangle+y_N\vert1\rangle+z_N\vert2\rangle)\rbrace\nonumber\\&&+\mathcal{A}_{33}\lbrace u_1\exp(i\varphi_1^{3})(u_N\vert0\rangle+v_N\vert1\rangle+w_N\vert2\rangle)\rbrace,\nonumber\\&&
\end{eqnarray}
where the coefficients $A_{ij}$ are calculated from
\begin{eqnarray}
\mathcal{A}=\prod\limits_{n=2}^{N}G_{n}\lbrace\varphi\rbrace,~~~\nonumber\\
G_{n}\lbrace\varphi\rbrace=\begin{pmatrix}\label{time dependent spin-1x}
    G_{11} & G_{12}& G_{13} \\
    G_{21}&G_{22} & G_{23}\\
   G_{31} & G_{32} & G_{33}
\end{pmatrix}.
\end{eqnarray} 
The matrix elements of Eq.~(\ref{time dependent spin-1}) is defined in Appendix~\ref{wavefunctionforspin-1}.

Again, for spin-1 case we study spin dynamics and analyze the Fourier power spectrum of operators defined as follows $I_{S_{x},S_{y},S_{z}}=\bigg\vert\int\limits_{-\infty}^{\infty}\langle S_{x,y,z}\rangle\exp(-i\omega t)dt\bigg\vert^{2}$, where $S_{x,y,z}$ are spin components for $S=1$ case. The Fourier power spectrum for $S_x$ and $S_z$ components
are plotted in Figs.~\ref{psdsigmaxspin1} for the regular $K=0.5$ and chaotic $K=10$ regimes. We clearly see from Figs.~\ref{psdsigmaxspin1} that in the  regular regime we get a few sharp peaks but in the chaotic regime we see broadening of the spectrum and many peaks which is a signature of chaos. The continuously filled lower band manifests the essence of the chaos in the spin dynamics.  Spin dynamics for $S_x$ and $S_z$ components for regular and chaotic cases are plotted in Figs.~\ref{regularyxspin1} (a)-(d). The spin dynamics clearly differentiates between regular and chaotic case. A quasi periodic oscillation is visible when the oscillator is in the regular regime and a chaotic oscillation for the oscillator in the chaotic regime.  Transition from the quasi periodic to the chaotic spin dynamics while changing the stochasticity parameter $K$ from $0.5$ to $10$ is a signature of chaos.
\begin{figure}[t!]
 \includegraphics[width=0.45\linewidth,height=2.4in]{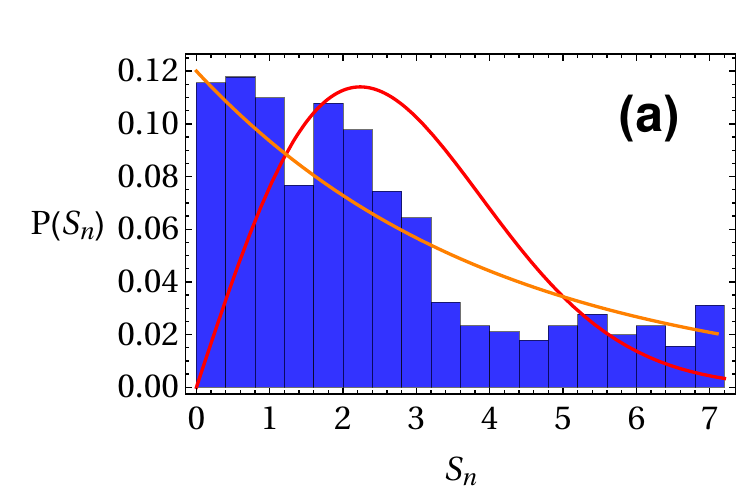}\ \includegraphics[width=0.45\linewidth,height=2.4in]{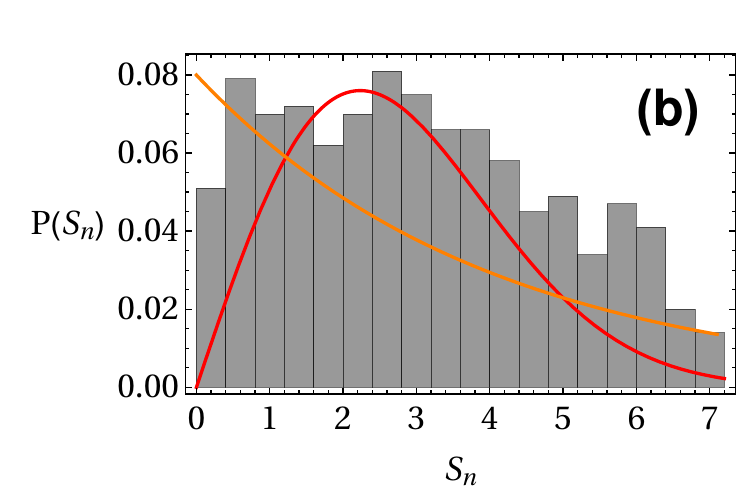}
\caption{Histogram plot of level statistics of Hamiltonian $\hat{H}_{n}=\hat{H}_{NV}+g\hat{V}_{c,NV}$ for spin-1 system  (a) in the regular regime at $K=0.5$ (Blue) and (b) in the chaotic regime $K=10$ (Gray). A reference plot for Poissonian statistics (Orange) and Gaussian statistics (Red) is also shown. $1000$ kicks are considered.
The parameters are $m=1$, $g=1$,  $\Omega=1$, $\delta=1$,  $\omega_{r}=0.2$, T=1, $\alpha=\pi/2$. The values of parameters: $K=\epsilon I_{0} T \frac{6\pi \mu}{m^2\omega_{0}^{2}}$, $\mu=\frac{\omega_{r}^{2}m}{2a_{0}^2}$ $I_{0}=\frac{m}{2}x_{0}^2 \omega_{r}$, $m=6\times10^{-17}$Kg, $x_{0}=a_{0}=5\times10^{-3}$m, T=10$\mu$s, $\omega_{r}=\omega_{0}=2\pi\times5\times10^{6}$Hz, for chaotic case $\epsilon=0.003$ and for the regular case $\epsilon=0.0003$.}
\label{levelstatisticsspin1}
 \end{figure} 
Following the recipes used for spin-1/2 case in section~\ref{spin_dynamics} we analyze the nearest-neighbour level statistics for spin-1 case. 
In the three-level system, two nearest-neighbour spacings at $n^{\rm th}$ kick are given as
\begin{eqnarray}\label{distance between levels of spin-1}
&& S_n^{1}=E_2^{(n)}-E_3^{(n)}=\frac{1}{2}\big(-\delta+\sqrt{(2\chi_{n}+\sqrt{2}\Omega)^2+\delta^2} \big),\nonumber\\
&& S_n^{2}=E_1^{(n)}-E_2^{(n)}=\frac{1}{2}\big(\delta+\sqrt{(2\chi_{n}+\sqrt{2}\Omega)^2+\delta^2} \big),
\end{eqnarray}
 We calculate the nearest-neighbour spacing for a few kicks and plot the distribution functions. For the calculation of  level-spacing distribution of the Hamiltonians at different kicks, we notice that the off-diagonal entries of the Hamiltonians contain $I_n$ and $\theta_n$ having range $[0,2\pi]$  with $(mean \sim 3.132$ and $variance\sim 3.382 )$ are stochastic. In the chaotic case of $S=1$, the distribution of the off-diagonal entries form a Gaussian ensemble with a mean of $0.49$ and a variance $5.7$ and level-spacing distribution is not the same as that of Gaussian orthogonal ensemble   \cite{haake1991quantum} but the effect of level repulsion is visible in this larger Hilbert space which was absent in the spin-1/2. In  Fig. \ref{levelstatisticsspin1} we show the level-spacing distribution of regular and chaotic regimes for $S=1$ case with a reference  Poissonian $P(S)\propto \exp{(-S)}$ and  Wigner-Dyson distributions $P(S)\propto(\pi S/2)\exp{(-\pi(S/2)^2)}$.  We see that the maxima of  distribution functions $P(S_n)$
in the regular case are shifted to the area of small $S_n$ and in the chaotic case to the finite $S_n$. Although distribution functions are not strictly Poissonian or Wigner-Dyson type, the effect of the level repulsion is attributed to the quantum chaotic phenomena is observed.

\section{ Statistical average over various $I_0$ and $\theta_0$ }\label{StatisticalSPIN1/2}
One of the principle differences between classical and quantum chaos is the sensitivity of the classical nonlinear dynamics with respect to the slight change of the initial conditions. Typically chaotic classical phase trajectories diverge in time when starting from the vicinity of the same region.
 \begin{figure}[H]
 \includegraphics[width=0.45\linewidth,height=2.2in]{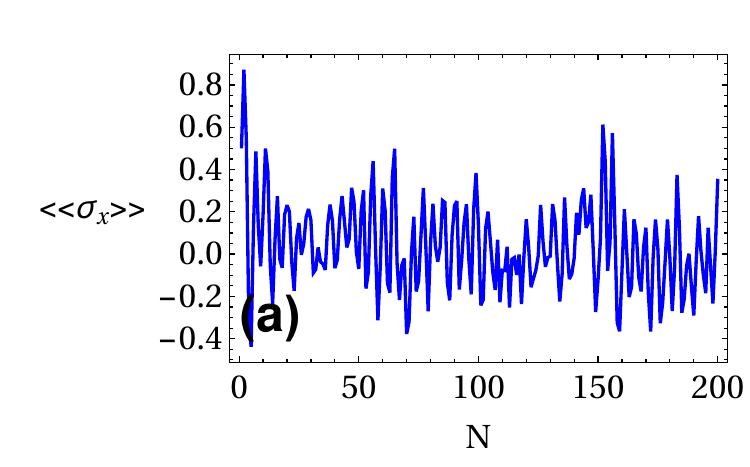}\ \includegraphics[width=0.45\linewidth,height=2.2in]{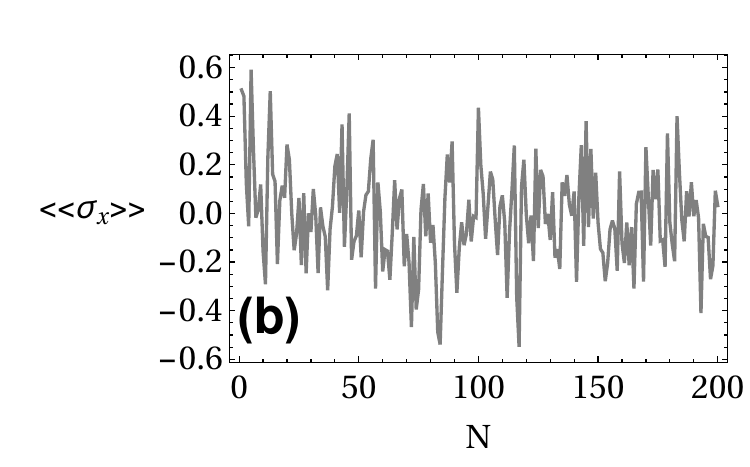}\\
 \includegraphics[width=0.45\linewidth,height=2.2in]{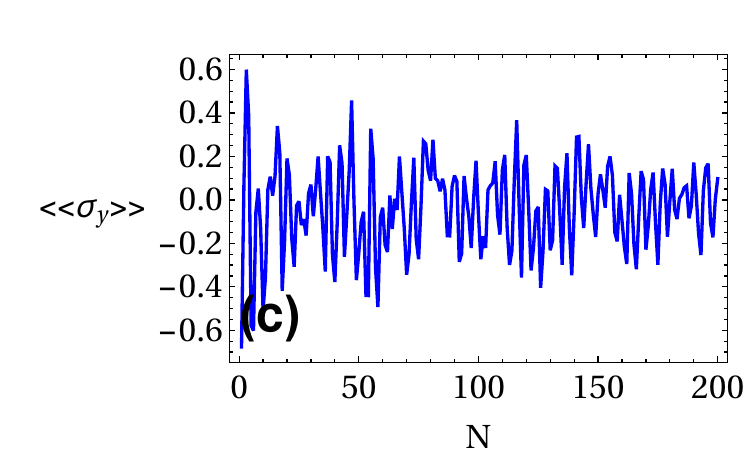}\ \includegraphics[width=0.45\linewidth,height=2.2in]{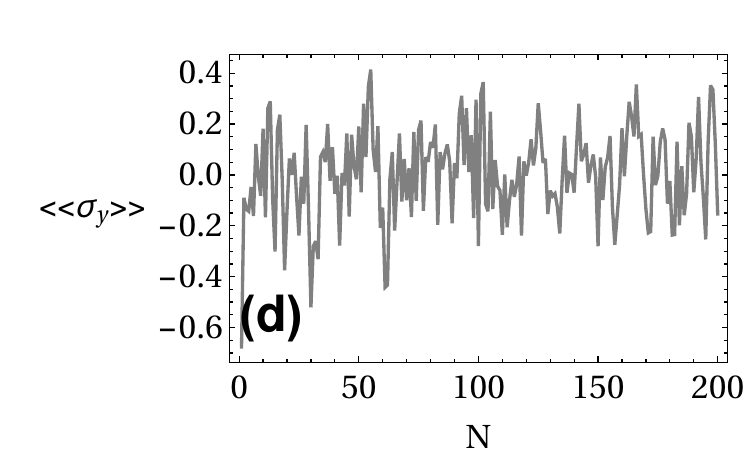}\\
 \includegraphics[width=0.45\linewidth,height=2.2in]{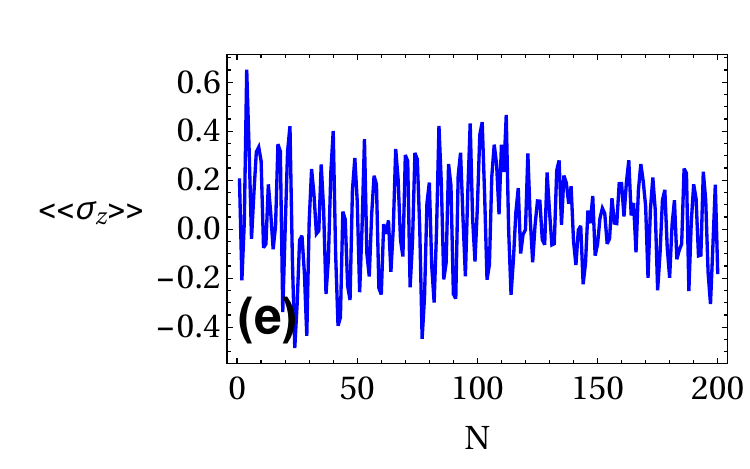}\ \includegraphics[width=0.45\linewidth,height=2.2in]{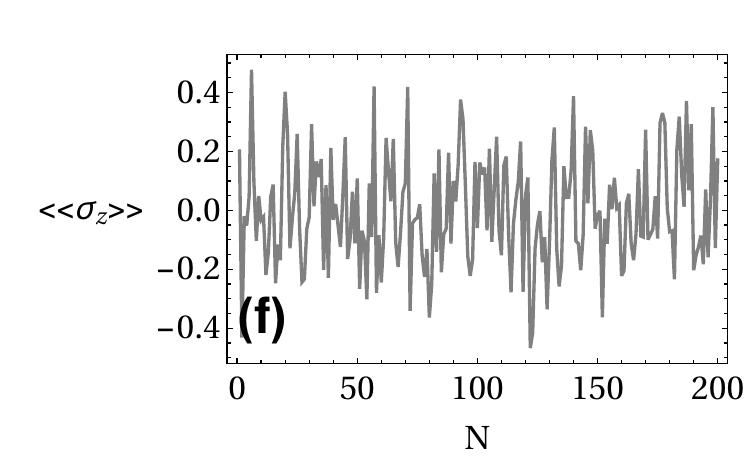}
\caption{Statistical average of spin dynamics (Spin-1/2 system) for  $\langle \langle \sigma_x\rangle \rangle$, $\langle \langle \sigma_y\rangle\rangle$  and $\langle\langle \sigma_z\rangle\rangle$  in the regular regime at $K=0.5$ ((a), (c) and (e)) and in the chaotic regime at $K=10$  ((b), (d) and (f)). For calculating statistical average of spin dynamics (Spin-1/2 system) we have taken 15 different sets of $(I_{0},\theta_{0})$.  The parameters are $m=1$, $g=1$,  $\omega_{0}=1$, $\omega_{r}=0.2$, $T=1$, $\alpha=\pi/2$. The values of the parameters in the real units: $K=\epsilon I_{0} T \frac{6\pi \mu}{m^2\omega_{r}^{2}}$, $\mu=\frac{\omega_{r}^{2}m}{2a_{0}^2}$, $I_{0}=\frac{m}{2}x_{0}^2 \omega_{r}$, $m=6\times10^{-17}$Kg, $x_{0}=a_{0}=5\times10^{-3}$m, $T=10\mu$s, $\omega_{r}=\omega_{0}=2\pi\times5\times10^{6}$Hz, for chaotic $\varepsilon=0.003$ and for regular $\varepsilon=0.0003$.}
\label{regularyxspin1x}
 \end{figure}
 \begin{figure}[H]
  \includegraphics[width=0.45\linewidth,height=2.2in]{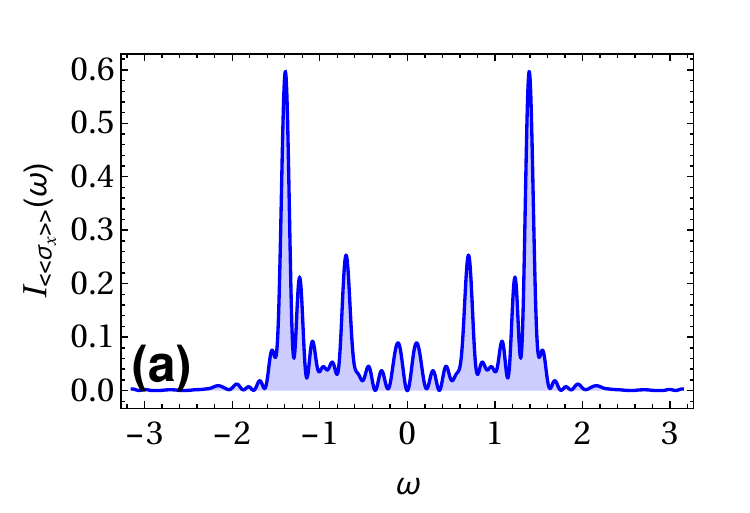}\ \includegraphics[width=0.45\linewidth,height=2.2in]{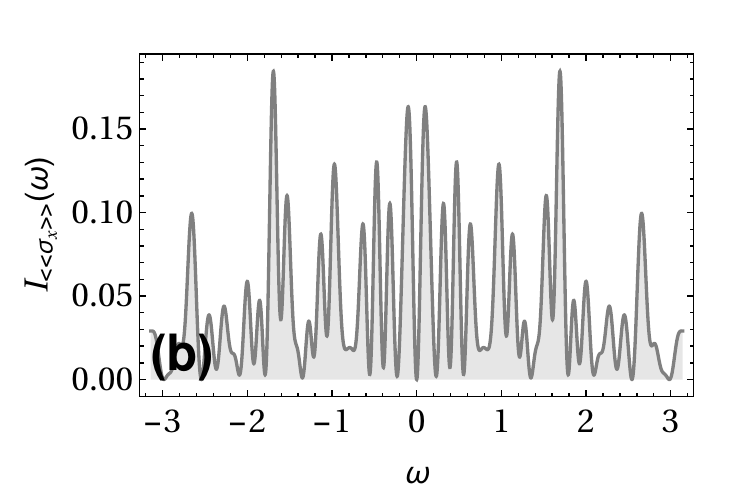}\\
 \includegraphics[width=0.45\linewidth,height=2.2in]{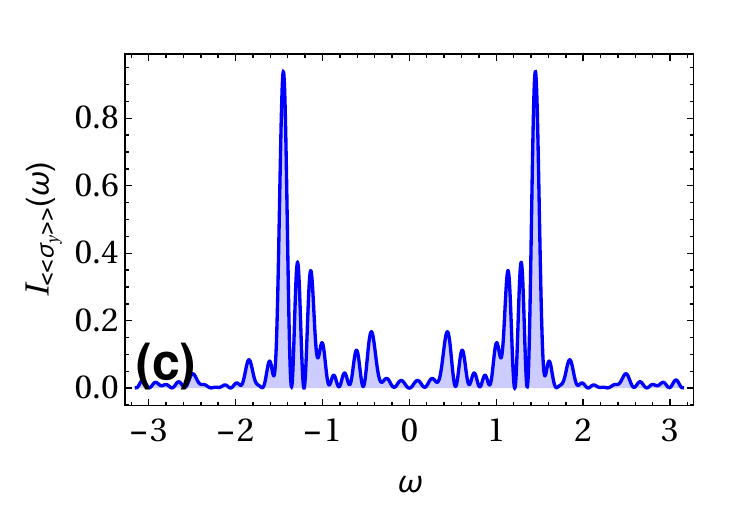}\ \includegraphics[width=0.45\linewidth,height=2.2in]{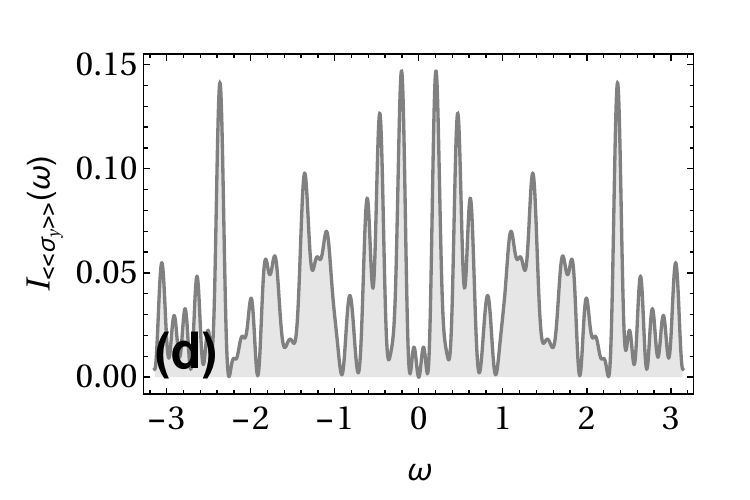}\\ \includegraphics[width=0.45\linewidth,height=2.2in]{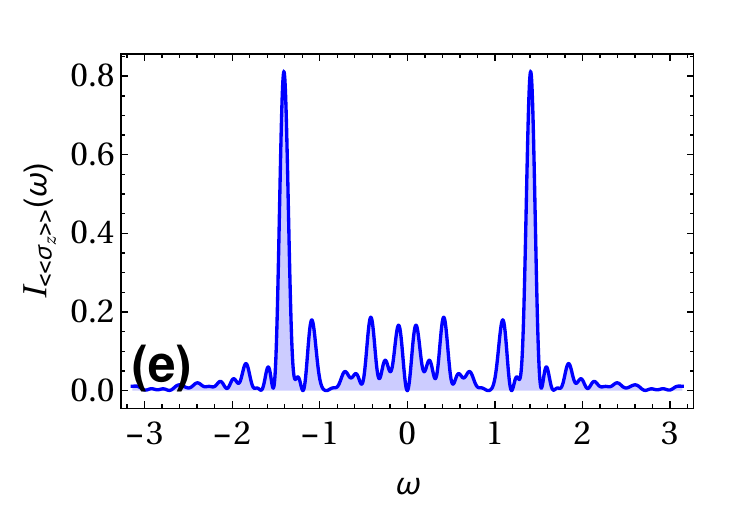}\ \includegraphics[width=0.45\linewidth,height=2.2in]{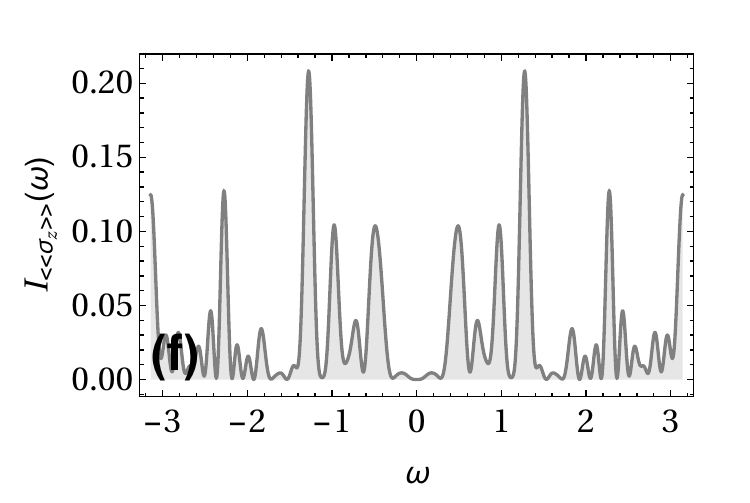}
\caption{Statistical average of Fourier Power spectrum density  (Spin-1/2 system) for  $\langle \langle \sigma_x\rangle \rangle$, $\langle \langle \sigma_y\rangle\rangle$  and $\langle\langle \sigma_z\rangle\rangle$ in the regular regime ((a), (c) and (e)) at
$K=0.5$ (Blue), and in the chaotic regime ((b), (d) and (f)) at $K=10$ (Gray). For calculating Statistical average of Spin dynamics(Spin-1/2 system) we have taken 15 different sets of $(I_{0},\theta_{0})$. The parameters used for the plot are $m=1$, $g=1$,  $\omega_{0}=1$, $\omega_{r}=0.2$, $T=1$, $\alpha=\pi/2$. The values of the parameters in the real units: $K=\epsilon I_{0} T \frac{6\pi \mu}{m^2\omega_{r}^{2}}$, $\mu=\frac{\omega_{r}^{2}m}{2a_{0}^2}$ $I_{0}=\frac{m}{2}x_{0}^2 \omega_{r}$, $m=6\times10^{-17}$Kg, $x_{0}=a_{0}=5\times10^{-3}$m, $T=10\mu$s, $\omega_{r}=\omega_{0}=2\pi\times5\times10^{6}$Hz, for chaotic case $\epsilon=0.003$ and for the regular case $\epsilon=0.0003$.}
\label{averagepsdsigmaxspin}
 \end{figure}
We want to know if classical chaos imposes certain 
effects on the quantum subsystem in the case of hybrid quantum-classical chaos. For this aim, we considered the statistical average over many initial values $I_0$ and $\theta_0$. 

Results of numeric calculations are presented in Figs.~\ref{regularyxspin1x} and Figs.~\ref{averagepsdsigmaxspin}.

As we see from the plots, quantum dynamics is less sensitive to the averaging performed on the classical part. Chaotic $K>1$ and regular $K<1$ characteristics of quantum dynamics is preserved after averaging done over the classical cantilever.

\section{Feedback Effect}\label{Feedback Effect}
An interesting question is the feedback of the quantum subsystem on the classical dynamics. 
For studying this problem one needs to solve recurrent relations self-consistently together with the Schr\"odinger equation. After transforming into the action-angle variables we deduce:

\begin{eqnarray}\label{semiclassical-1}
&& \frac{d|\psi\rangle}{dt}=-\frac{i}{\hbar}\left(\hat H_s+g\sqrt{2I/m\omega_r}\cos\theta \hat S_z\right)|\psi\rangle,\nonumber\\
  &&\frac{dI}{dt}=-\frac{\partial H_{I,\theta}}{\partial\theta}-\frac{\partial \hat V_{c,NV}}{\partial\theta}=g\sqrt{2I/m\omega_r}\sin\theta \langle\psi|\hat{S}_z|\psi\rangle\nonumber\\&&-\varepsilon \frac{\partial V(I,\theta)}{\partial \theta}T\sum\limits_{n=-\infty}^{\infty}\delta\left(t-nT\right), \nonumber\\
  &&\frac{d\theta}{dt}=\frac{\partial H_{I,\theta}}{\partial I}-\frac{\partial \hat V_{c,NV}}{\partial I}=-\frac{g}{\sqrt{2m\omega_{r}I}}\cos\theta\langle\psi|\hat{S}_z|\psi\rangle+\omega(I)+\nonumber\\&&\varepsilon \frac{\partial V(I,\theta)}{\partial I}T\sum\limits_{n=-\infty}^{\infty}\delta\left(t-nT\right).\nonumber\\
\end{eqnarray}
The standard procedure for solving Eq.(\ref{semiclassical-1}) consists of two steps: free propagation and kick. During the free propagation, the effect of kicks is absent and vice versa. We note that our system is inherently nonlinear, and nonlinearity is a part of the main Hamiltonian. The nonlinearity in our case is not weak, and the model is non-perturbative. While action $I$ is an adiabatic variable, angle $\theta$ is a fast oscillating variable such that $T\theta> 2\pi$. The formal solution of the recurrent relations has a form of morphism $\mathcal{M}=I_{n},\theta_n\rightarrow I_{n+1},\theta_{n+1}$, where $n, n+1$ corresponds to the values after $n$th and $(n+1)^{\rm th}$ kick, respectively. We have two time scales in the problem, fast and slow.
\begin{figure}[t]
\includegraphics[width=0.45\linewidth,height=2.2in]{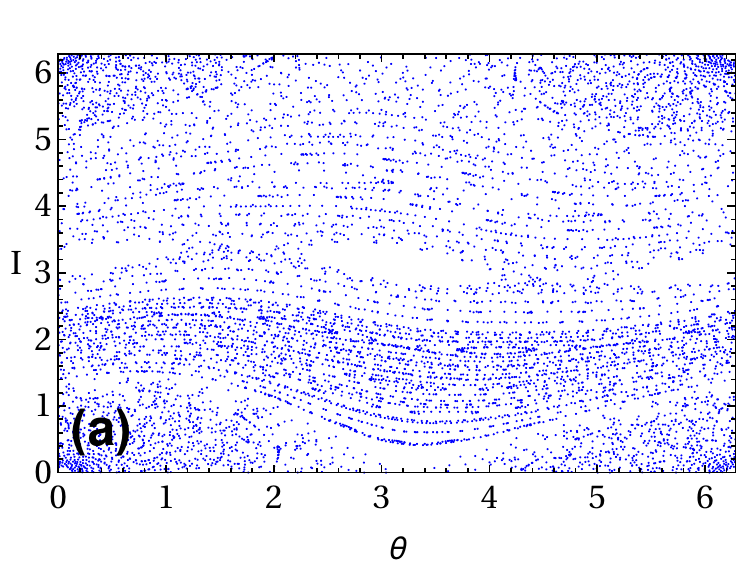}\ \includegraphics[width=0.45\linewidth,height=2.2in]{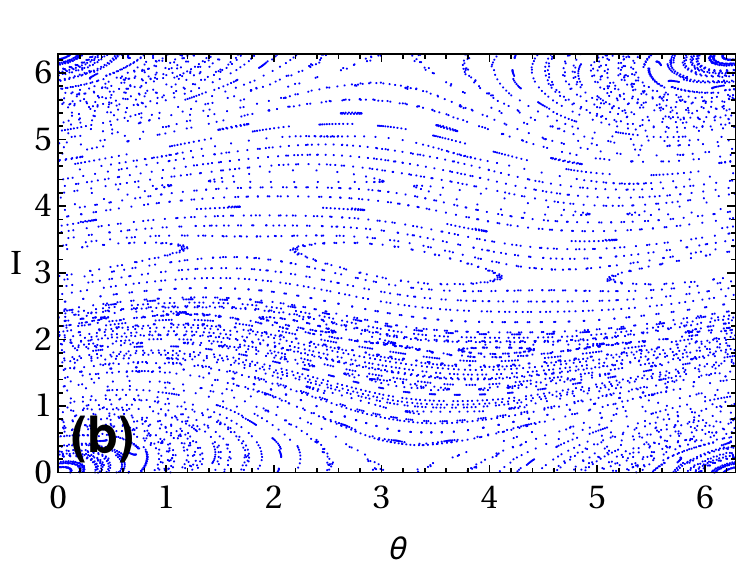}
\caption{The Phase space plot of cantilever's dynamics constructed through the recurrence relations Eq. (\ref{semiclassica19}) with feedback in ((a) and (b)) the regular regime $K=0.5$ (Blue) where the phase space is covered by two different phase trajectories: open hyperbolic and some part of closed ellipse.The parameters for fig.\ref{regularfeed}(a) are $m=1$, $g=0.1$,  $\omega_{0}=10$, $\omega_{r}=0.2$, $T=1$. The parameters for fig.\ref{regularfeed}(b) are $m=1$, $g=0.01$,  $\omega_{0}=1$, $\omega_{r}=0.2$, $T=1$. The values of the parameters in the real units: $K=\epsilon I_{0} T \frac{6\pi \mu}{m^2\omega_{r}^{2}}$, $\mu=\frac{\omega_{r}^{2}m}{2a_{0}^2}$ $I_{0}=\frac{m}{2}x_{0}^2 \omega_{r}$, $m=6\times10^{-17}$Kg,
$x_{0}=a_{0}=5\times10^{-3}$m, $T=10\mu$s, $\omega_{r}=\omega_{0}=2\pi\times5\times10^{6}$Hz, for the regular case $\varepsilon=0.0003$.}
\label{regularfeed}
 \end{figure}
The time unit for the evolution of $I$ and $\theta$ is T. Meaning that on the times shorter than $t<T$ variables $I_n$, $\theta_n$ are constants. To go from $I_n$, $\theta_n$ to $I_{n+1}$, $\theta_{n+1}$ we need at least time $t=T$. On the other hand we have fast time oscillations in the Schr\"odinger equation because $\omega_0 \gg g$. However, these fast phase oscillations of the wave function are distinct from the
evolution of the wave function that occurs on the larger time scale $t>T$  due to the evolution of $I_n$, $\theta_n$. Existence of fast and slow time scales in the system allows us to tackle the feedback problem in the following scheme: In order to obtain fast time evolution of the wave function valid for $t<T$, we solve the first equation in Eq.(\ref{semiclassical-1}) for a constant $I_n$, $\theta_n$ (for $t<T$, variables $I_{n}$ and $\theta_{n}$ are constant). We solve Schr\"odinger equation analytically:
\begin{eqnarray}\label{semiclassical-x}
&& \frac{d|\psi\rangle}{dt}=-\frac{i}{\hbar}\left(\hat H_s+g\sqrt{2I/m\omega_r}\cos\theta \hat S_z\right)|\psi\rangle,
\end{eqnarray}
where $\hat{S_{z}}=\frac{1}{2}(\cos{\alpha}\sigma_{z}+\sin{\alpha}(\sigma^{+}+\sigma^{-}))$. When $I_{n}$ and $\theta_{n}$ are constants, $a_{n}=\sqrt{2I_{n}/m\omega_r}\cos\theta_{n}=V_{0}(I_{n})\cos\theta_{n}$ is also a constant.

After solving Schr\"odinger's equation analytically, we get the evolved wave function as:
\begin{eqnarray}
\vert\psi\rangle=\begin{pmatrix}
\Xi_{1}  \\
\Xi_{2} 
\end{pmatrix},
\end{eqnarray}
where
\begin{eqnarray}
\Xi_{1}=\frac{-i a_{n} g\sin{\frac{1}{2}t\sqrt{a_{n}^2g^2+\omega_{0}^2}}}{\sqrt{a_{n}^2g^2+\omega_{0}^2}}, 
\end{eqnarray}
and
\begin{eqnarray}
\Xi_{2}=\cos{\frac{1}{2}t\sqrt{a_{n}^2g^2+\omega_{0}^2}}+\frac{i \omega_{0}\sin{\frac{1}{2}t\sqrt{a_{n}^2g^2+\omega_{0}^2}}}{\sqrt{a_{n}^2g^2+\omega_{0}^2}}. 
\end{eqnarray}
 Here we introduced shorthand notation $\Omega_{n}=\sqrt{a_{n}^2g^2+\omega_{0}^2}$. We note that $\omega_{0}$ is a large parameter of the proposed theoretical model and this assumption is based on the value of $\omega_{0}=2.88$GHz for NV centres. Therefore $\omega_{0}\gg a_{n}g$ and $\sqrt{a_{n}^2g^2+\omega_{0}^2}\approx\omega_{0}$. 
 \begin{figure}[H]
  \includegraphics[width=0.45\linewidth,height=2.2in]{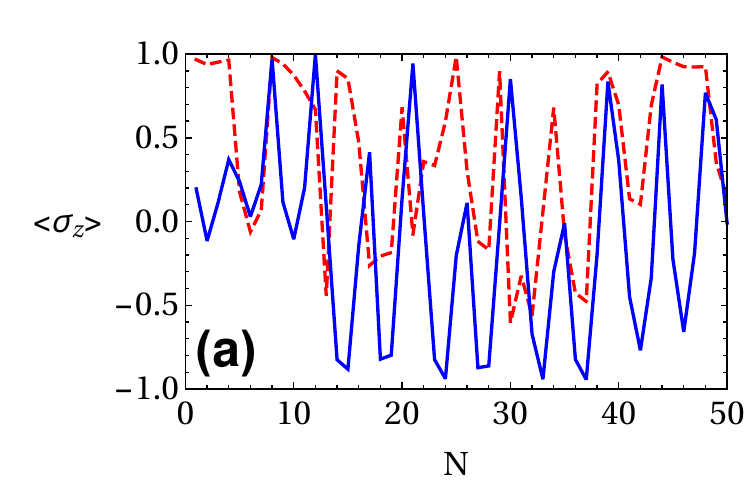}\ \includegraphics[width=0.45\linewidth,height=2.2in]{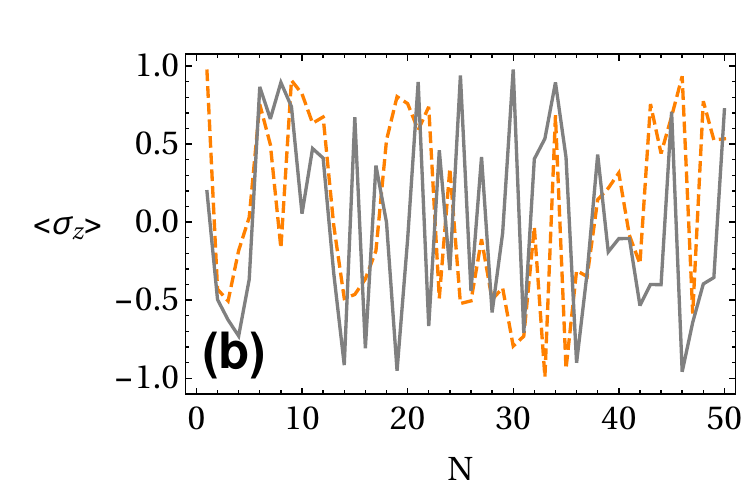}
\caption{Spin dynamics with feedback (Solid) and without feedback (Dashed)  for $\langle\sigma_z\rangle$  in the regular regime at $K=0.5$ (see (a)) and $\langle\sigma_z\rangle$ in the chaotic regime at $K=10$ (see (b)). The parameters are $m=1$, $g=1$,  $\omega_{0}=1$,
$\omega_{r}=0.2$, $T=1$, $\alpha=\pi/2$. The values of the parameters in the real units: $K=\epsilon I_{0} T \frac{6\pi \mu}{m^2\omega_{r}^{2}}$, $\mu=\frac{\omega_{r}^{2}m}{2a_{0}^2}$, $I_{0}=\frac{m}{2}x_{0}^2 \omega_{r}$, $m=6\times10^{-17}$Kg, $x_{0}=a_{0}=5\times10^{-3}$m, $T=10\mu$s, $\omega_{r}=\omega_{0}=2\pi\times5\times10^{6}$Hz, for chaotic $\varepsilon=0.003$ and for regular $\varepsilon=0.0003$.}
\label{regularyx1}
 \end{figure}
 To obtain the feedback term in the explicit form we calculate $\langle\psi(t)\vert\hat{S_{z}}\vert\psi(t)\rangle$ and deduce:
 \begin{eqnarray}\label{semiclassica19xx}
\int_{0}^{T}\langle\psi(t)\vert\hat{S_{z}}\vert\psi(t)\rangle dt = \frac{-a_{n}g\omega_{0}T}{2\Omega_{n}^2}=\frac{-V_{0}(I_{n})\cos{\theta_{n}}gT}{2\omega_{0}}.
\end{eqnarray}
Consequently Eq.(\ref{semiclassical-1}) takes the form:
\begin{eqnarray}\label{semiclassica17}
&&I_{n+1}=I_{n}+g\sqrt{2I_{n}/m\omega_r}\sin\theta_{n}\int\limits_{0}^{T}\langle\psi(t)|\hat{S}_z|\psi(t)\rangle dt-K \sin{\theta},\nonumber\\
&&\theta_{n+1}=\theta_{n}+I_{n+1}-g\frac{\cos\theta_{n}}{\sqrt{2m\omega_rI_{n}}}\int\limits_0^T\langle\psi(t)|\hat{S}_z|\psi(t)\rangle dt.
\end{eqnarray}
The explicit integrated feedback term Eq.(\ref{semiclassica19xx}) is plugged in the Eq.({\ref{semiclassica17}}) and the generalized standard map is deduced in the form:
\begin{eqnarray}\label{semiclassica19}
&&I_{n+1}=I_{n}-\frac{g^2TV_{0}^{2}(I_{n})}{4\omega_0}\sin2\theta_{n}-K \sin{\theta},\nonumber\\
&&\theta_{n+1}=\theta_{n}+I_{n+1}+g^2T\frac{\cos^{2}\theta_{n}}{2m\omega_r\omega_{0}}.
\end{eqnarray}
In Fig.~\ref{regularfeed} dynamics of cantilever with feedback effects in regular regime for $K=0.5$ is shown. We see two cases: $g=0.1$ and $g=0.01$. When the interaction strength between NV spin and cantilever is moderate ($g=0.1$), we see in Fig.~\ref{regularfeed}(a) a small deviation from regular dynamics in presence of feedback. For small interaction between NV spin and cantilever, as shown in  Fig.~\ref{regularfeed}(b), feedback does not effect the dynamics of cantilever. In Fig.~\ref{regularyx1}, we compare the spin dynamics with and without feedback effects. We see a minor change in the amplitude of oscillations in regular and chaotic cases due to the feedback term. In case of the regular regime, the feedback not much affect the magnetization as compared to the dynamics without feedback term.  Similarly, the switching pattern is hardly affected in the chaotic regime.

\section{Conclusions}
\label{summary}
In the present work, we studied hybrid quantum-classical NEMS  systems. The classical part comprised of a nanocantilever, and the quantum part is the NV spin. Nanocantilever performs nonlinear oscillations in the chaotic and regular regimes. Due to the spin-cantilever coupling, the effects of the oscillations of the cantilever are transmitted to the spin dynamics. The problem in question was whether the classical dynamical chaos may induce quantum chaos or other effects of quantum stochasticity in the quantum dynamics of the NV spin. We studied the Poincar\'{e} section of spin-dynamics and explored the Fourier power spectral density of the quantum dynamical observables in the chaotic and regular regimes. We investigated the generation of quantum coherence for the NV center coupled to nanocantilever in the chaotic and regular regime. We also investigated the quantum Poincar\'{e} recurrence in the chaotic and regular regime.  While the Fourier spectrum analysis clearly indicates the presence of stochasticity in the dynamics of quantum observables, some characteristics of quantum chaos are absent. The dynamical chaos imposed to the cantilever dynamics through the kicking induces the stochastic dynamics on the quantum subsystem. However, this stochastic dynamics of the classical cantilever does not manifest all the features of quantum chaos. We also investigated a three-level system for the quantum part considering NV spin as spin-1 particle. We see that the Fourier power spectrum and spin dynamics evince the effects of chaos. For spin-1 case we see a quasi-Gaussian distribution of nearest-neighbour level spacing for the oscillator in chaotic regime and quasi-Poissonian level statistics for the oscillator in regular regime. We also explore the effect of quantum feedback on classical cantilever in both cases regular and chaotic and also see the effect on spin dynamics. Feedback effect is negligible  in the chaotic regime of the system.

\vspace{0.5cm}
{\bf Author contribution statement:} \\
\noindent
AKS, LC, ZT, IT and SKM conceived of the presented idea and developed the theory and performed analysis. AKS performed numerical calculations. All authors discussed the results and contributed to the final manuscript.

\vspace{0.5cm}
{\bf Acknowledgement:}\\
\noindent
AKS would like to thank 
 D.V. Khomitsky and Rohit Kumar Shukla for his valuable suggestions.

\appendix
\section*{Appendix}
\section{Normalization condition for the wave function represented in Eq.(\ref{time dependent}) when $G_{n}\lbrace\varphi\rbrace$ is diagonal. }
\label{appendix1}
The eigenstates and eigenvalues of the $n$th Floquet operator have the form:
\begin{eqnarray}\label{the explicit form1}
&&\vert\varphi_{n}^+\rangle=\eta_n\vert0\rangle+\xi_n\vert1\rangle, \nonumber\\
&&\vert\varphi_{n}^-\rangle=\xi_n\vert0\rangle-\eta_n\vert1\rangle,
\end{eqnarray}
where
\begin{eqnarray}\label{eigenphase1}
&&\eta_n=\frac{1}{\sqrt{1+k_n^2}}, \nonumber\\
&&\xi_n=-\frac{k_n}{\sqrt{1+k_n^2}}, \nonumber\\&& k_n=\frac{\omega_{0}+\sqrt{\chi_{n}^2+\omega_{0}^2}}{\chi_{n}}.
\end{eqnarray}
 $k_n=k_n^{*}$, so $\eta_n=\eta_n^{*}$ and $\xi_n=\xi_n^{*}$.
 The normalization condition of the wave function after $N$ kickes, i. e. , at $t=NT$ has the form:
 
\begin{eqnarray*}\label{Normalization1}
&&\langle\psi(t=TN)\vert\psi(t=TN)\rangle=\vert A_{11}\vert^{2}\lbrace \vert\eta_{1}\vert^2(\vert\eta_{N}\vert^2+\vert\xi_{N}\vert^2)
\rbrace+A_{11}^{*}A_{12}\lbrace \vert\eta_{1}\vert^2(\xi_{N}\eta_{N}^{*}\nonumber\\&&-\xi_{N}^{*}\eta_{N}) \rbrace+A_{11}^{*}A_{21} \lbrace \eta_{1}^{*}\xi_{1}\exp(2 i \varphi_{1})(\vert\eta_{N}\vert^2+\vert\xi_{N}\vert^2)\rbrace\nonumber\\&& +A_{11}^{*}A_{22}\lbrace \eta_{1}^{*}\xi_{1}(\xi_{N}\eta_{N}^{*}-\xi_{N}^{*}\eta_{N}) \rbrace+A_{11}A_{12}^{*}\lbrace \vert\eta_{1}\vert^2(-\xi_{N}\eta_{N}^{*}+\xi_{N}^{*}\eta_{N}) \rbrace\nonumber\\&& +\vert A_{12}\vert^{2}\lbrace \vert\eta_{1}\vert^2(\vert\eta_{N}\vert^2+\vert\xi_{N}\vert^2) \rbrace+A_{12}^{*}A_{21}\lbrace \eta_{1}^{*}\xi_{1}\exp(2i\varphi_{1})(-\xi_{N}\eta_{N}^{*}+\xi_{N}^{*}\eta_{N}) \rbrace\nonumber\\&& +A_{12}^{*}A_{22} \lbrace \eta_{1}^{*}\xi_{1}\exp(2 i \varphi_{1})(\vert\xi_{N}\vert^2+\vert\eta_{N}\vert^2)\rbrace+A_{21}^{*}A_{11} \lbrace \eta_{1}\xi_{1}^{*}\exp(-2 i \varphi_{1})(\vert\xi_{N}\vert^2\nonumber\\&& +\vert\eta_{N}\vert^2)\rbrace +A_{21}^{*}A_{12} \lbrace \eta_{1}\xi_{1}^{*}\exp(-2 i \varphi_{1})(\xi_{N}\eta_{N}^{*}-\eta_{N}\xi_{N}^{*})\rbrace\nonumber\\&&+\vert A_{21}\vert^{2}\lbrace \vert\xi_{1}\vert^2(\vert\eta_{N}\vert^2+\vert\xi_{N}\vert^2) \rbrace+A_{21}^{*}A_{22} \lbrace \vert\xi_{1}\vert^{2}(\eta_{N}^{*}\xi_{N}-\xi_{N}^{*}\eta_{N})\rbrace\nonumber\\&&+A_{11}A_{22}^{*}\lbrace\eta_{1}\xi_{1}^{*}\exp(-2i\varphi_{1})(\eta_{N}\xi_{N}^{*}-\eta_{N}^{*}\xi_{N})\rbrace+A_{12}A_{22}^{*} \lbrace \eta_{1}\xi_{1}^{*}\exp(-2
i \varphi_{1})(\vert\xi_{N}\vert^{2}+\vert\eta_{N}\vert^2)\rbrace\nonumber\\&&+A_{21}A_{22}^{*}\lbrace\vert\xi_{1}\vert^2(\eta_{N}\xi_{N}^{*}-\eta_{N}^{*}\xi_{N})+\vert A_{22}\vert^{2}\lbrace \vert\xi_{1}\vert^2(\vert\eta_{N}\vert^2+\vert\xi_{N}\vert^2) \rbrace.
\end{eqnarray*}
In the particular case, $A_{11}=A_{22}^{*}$ and $A_{12}=A_{21}^{*}$\\
after simplification one can get the form:
\begin{eqnarray}\label{Normalization2}
&&\langle\psi(t=TN)\vert\psi(t=TN)\rangle=\vert A_{11}\vert^{2}+\vert A_{12}\vert^{2}+A_{11}^{*}A_{21} \lbrace \eta_{1}^{*}\xi_{1}\exp(2 i \varphi_{1})\rbrace\nonumber\\&&+A_{12}^{*}A_{22} \lbrace \eta_{1}^{*}\xi_{1}\exp(2 i \varphi_{1})\rbrace+A_{21}^{*}A_{11} \lbrace \eta_{1}\xi_{1}^{*}\exp(-2 i \varphi_{1})\rbrace+A_{12}A_{22}^{*} \lbrace \eta_{1}\xi_{1}^{*}\exp(2 i \varphi_{1})\rbrace.\nonumber\\
\end{eqnarray}
If all kicks are identical, the off-diagonal elements of the matrix $\mathcal{A}$ are zero. Therefore:
$\vert A_{11}\vert^{2}=1$, $\vert A_{12}\vert^{2}=0$, $\vert A_{21}\vert^{2}=0$, $\vert A_{22}\vert^{2}=1$\\
\begin{eqnarray}
&&\langle\psi(t=TN)\vert\psi(t=TN)\rangle=1.
\end{eqnarray}
\section{Normalization condition for the wave function represented in Eq.(\ref{time dependent}) when $G_{n}\lbrace\varphi\rbrace$ is non-diagonal. }
\label{Normalizationx}
We prove the normalization of the evolved wave function in the general case of three $N=3$ different kicks. The elements of the $G$ matrix in Eq.(\ref{time dependentx}) have the form:
\begin{eqnarray}
G_2\lbrace\varphi\rbrace=\begin{bmatrix}
    \exp(-i\varphi_2)(\eta_2\eta_{1}+\xi_2\xi_{1}) & \exp(-i\varphi_2)(\eta_2\xi_{1}-\xi_2\eta_{1}) \\
    \exp(i\varphi_2)(\xi_2\eta_{1}-\eta_2\xi_{1}) & \exp(i\varphi_2)(\eta_2\eta_{1}+\xi_2\xi_{1})
\end{bmatrix},
\end{eqnarray}
\begin{eqnarray}
G_3\lbrace\varphi\rbrace=\begin{bmatrix}
    \exp(-i\varphi_3)(\eta_3\eta_{2}+\xi_3\xi_{2}) & \exp(-i\varphi_3)(\eta_3\xi_{2}-\xi_3\eta_{2}) \\
    \exp(i\varphi_3)(\xi_3\eta_{2}-\eta_3\xi_{2}) &
    \exp(i\varphi_3)(\eta_3\eta_{2}+\xi_3\xi_{2})
\end{bmatrix}.
\end{eqnarray}
Therefore for $\mathcal{A}$ matrix we deduce
\begin{eqnarray}
&&\mathcal{A}=\prod\limits_{n=2}^{3}G_n\lbrace\varphi\rbrace,
\end{eqnarray}
\begin{eqnarray}
&&A_{11}=\exp(-i \varphi_{2}-i \varphi_{3})(\eta_{2}\eta_{1}+\xi_{2}\xi_{1})(\eta_{3}\eta_{2}+\xi_{3}\xi_{2})\nonumber\\&&+\exp(-i \varphi_{2}+i \varphi_{3})(\eta_{2}\xi_{1}-\xi_{2}\eta_{1})(\xi_{3}\eta_{2}-\eta_{3}\xi_{2}),
\end{eqnarray}
\begin{eqnarray}
&&A_{12}=\exp(-i \varphi_{2}-i \varphi_{3})(\eta_{2}\eta_{1}+\xi_{2}\xi_{1})(\eta_{3}\xi_{2}-\xi_{3}\eta_{2})\nonumber\\&&+\exp(-i \varphi_{2}+i \varphi_{3})(\eta_{2}\xi_{1}-\xi_{2}\eta_{1})(\eta_{3}\eta_{2}+\xi_{3}\xi_{2}),
\end{eqnarray}
\begin{eqnarray}
&&A_{21}=\exp(i \varphi_{2}-i\varphi_{3})(\eta_{3}\eta_{2}+\xi_{2}\xi_{3})(\eta_{1}\xi_{2}-\xi_{1}\eta_{2})\nonumber\\&&+\exp(i \varphi_{2}+i \varphi_{3})(\eta_{2}\eta_{1}+\xi_{2}\xi_{1})(\xi_{3}\eta_{2}-\eta_{3}\xi_{2}),
\end{eqnarray}
\begin{eqnarray}
&&A_{22}=\exp(i \varphi_{2}-i \varphi_{3})(\xi_{2}\eta_{1}+\eta_{2}\xi_{1})(\eta_{3}\xi_{2}-\xi_{3}\eta_{2})\nonumber\\&&+\exp(i \varphi_{2}+i \varphi_{3})(\eta_{2}\eta_{1}+\xi_{2}\xi_{1})(\eta_{3}\eta_{2}+\xi_{3}\xi_{2}),
\end{eqnarray}
\begin{eqnarray}\label{Normalization non diagonal}
\nonumber\\&&\langle\psi(t=TN)\vert\psi(t=TN)\rangle=\vert A_{11}\vert^{2}\vert\eta_{1}\vert^{2}+\vert A_{12}\vert^{2}\vert\eta_{1}\vert^{2}\nonumber\\&&+\vert A_{21}\vert^{2}\vert\xi_{1}\vert^2+A_{11}^{*}A_{21} \lbrace \eta_{1}^{*}\xi_{1}\exp(2 i \varphi_{1})\rbrace+A_{12}^{*}A_{22} \lbrace \eta_{1}^{*}\xi_{1}\exp(2 i \varphi_{1})\rbrace\nonumber\\&&+\vert A_{22}\vert^{2}\vert\xi_{1}\vert^2+A_{21}^{*}A_{11} \lbrace \eta_{1}\xi_{1}^{*}\exp(-2 i \varphi_{1})\rbrace\nonumber\\&&+A_{12}A_{22}^{*} \lbrace \eta_{1}\xi_{1}^{*}\exp(2 i \varphi_{1})\rbrace,
\end{eqnarray}
\begin{eqnarray}
&&\vert A_{11}\vert^2=((\eta_2 \xi_1 - \eta_1 \xi_2) (-\eta_3 \xi_2 + \eta_2 \xi_3) \cos[\varphi_2 - \varphi_3] + (\eta_1 \eta_2 + \xi_1 \xi_2) (\eta_2
\eta_3 \nonumber\\&&+ \xi_2 \xi_3) \cos[\varphi_2 + \varphi_3])^2 + ((\eta_2 \xi_1
- \eta_1 \xi_2) (-\eta_3 \xi_2 + \eta_2 \xi_3) \sin[\varphi_2 -
\varphi_3] \nonumber\\&&+ (\eta_1 \eta_2 + \xi_1 \xi_2) (\eta_2 \eta_3 +
\xi_2 \xi_3) \sin[\varphi_2 + \varphi_3])^2,
\end{eqnarray}
\begin{eqnarray}
&&\vert A_{12}\vert^2=((\eta_2 \xi_1 - \eta_1 \xi_2) (\eta_2 \eta_3 + \xi_2 \xi_3)
\cos[\varphi_2 - \varphi_3] + (\eta_1 \eta_2 + \xi_1 \xi_2) (\eta_3 \
\xi_2 \nonumber\\&&- \eta_2 \xi_3) \cos[\varphi_2 + \varphi_3])^2 + ((\eta_2 \xi_1 \
- \eta_1 \xi_2) (\eta_2 \eta_3 + \xi_2 \xi_3) \sin[\varphi_2 - \
\varphi_3] \nonumber\\&&- (\eta_1 \eta_2 + \xi_1 \xi_2) (-\eta_3 \xi_2 + \
\eta_2 \xi_3) \sin[\varphi_2 + \varphi_3])^2,
\end{eqnarray}
\begin{eqnarray}
&&\vert A_{21}\vert^2=((-\eta_2 \xi_1 + \eta_1 \xi_2) (\eta_2 \eta_3 + \xi_2 \xi_3)
\cos[\varphi_2 - \varphi_3] + (\eta_1 \eta_2 + \xi_1 \xi_2) (-\eta_3
\xi_2 \nonumber\\&&+ \eta_2 \xi_3) \cos[\varphi_2 + \varphi_3])^2 + ((-\eta_2
\xi_1 + \eta_1 \xi_2) (\eta_2 \eta_3 + \xi_2 \xi_3) \
\sin[\varphi_2 - \varphi_3] \nonumber\\&&+ (\eta_1 \eta_2 + \xi_1 \xi_2) (-\eta_3 \
\xi_2 + \eta_2 \xi_3) \sin[\varphi_2 + \varphi_3])^2,
\end{eqnarray}
\begin{eqnarray}
&&\vert A_{22}\vert^2=((\eta_2 \xi_1 - \eta_1 \xi_2) (-\eta_3 \xi_2 + \eta_2 \xi_3) \
\cos[\varphi_2 - \varphi_3] + (\eta_1 \eta_2 + \xi_1 \xi_2) (\eta_2
\eta_3 \nonumber\\&&+ \xi_2 \xi_3) \cos[\varphi_2 + \varphi_3])^2 + ((\eta_2 \xi_1
- \eta_1 \xi_2) (-\eta_3 \xi_2 + \eta_2 \xi_3) \sin[\varphi_2 -
\varphi_3] \nonumber\\&&+ (\eta_1 \eta_2 + \xi_1 \xi_2) (\eta_2 \eta_3 +
\xi_2 \xi_3) \sin[\varphi_2 + \varphi_3])^2,
\end{eqnarray}
\begin{eqnarray}
&& A_{21}A_{11}^{*}=-\exp(2 i \varphi_2) ((\eta_2^2 + \xi_2^2) (\eta_3 \xi_1 - \eta_1
\xi_3) \cos[\varphi_3] -
  i (\eta_2^2 (\eta_3 \xi_1 + \eta_1 \xi_3) \nonumber\\&& - \xi_2^2 (\eta_3
\xi_1+ \eta_1 \xi_3) +
      2 \eta_2 \xi_2 (-\eta_1 \eta_3 + \xi_1 \xi_3))
\sin[\varphi_3]) ((\eta_2^2 + \xi_2^2) (\eta_1 \eta_3 \nonumber\\&&+ \xi_1
\xi_3) \cos[\varphi_3] +
   i (\eta_1 (\eta_2^2 \eta_3 - \eta_3 \xi_2^2 +
         2 \eta_2 \xi_2 \xi_3) + \xi_1 (2 \eta_2 \eta_3 \xi_2
- \eta_2^2 \xi_3 \nonumber\\&&+ \xi_2^2 \xi_3)) \sin[\varphi_3]),
\end{eqnarray}
\begin{eqnarray}
&& A_{21}^{*}A_{11}=\exp(-2 i \varphi_2) ((-\eta_2 \xi_1 + \eta_1 \xi_2) (\eta_1 \eta_2 \
+ \xi_1 \xi_2) (\xi_2 (-\eta_3 + \xi_3) \nonumber\\&&+ \eta_2 (\eta_3 +
\xi_3)) (\eta_2 (\eta_3 - \xi_3) + \xi_2 (\eta_3 + \xi_3)) + (\
\eta_3 \xi_2 - \eta_2 \xi_3) (\eta_2 \eta_3 + \xi_2 \xi_3)
\nonumber\\&&(-(\xi_1 (-\eta_2 + \xi_2) + \eta_1 (\eta_2 + \xi_2)) (\eta_1
(\eta_2 - \xi_2) + \xi_1 (\eta_2 + \xi_2)) \cos[2 \varphi_3] \nonumber\\&&+
      i (\eta_1^2 + \xi_1^2) (\eta_2^2 + \xi_2^2) \sin[2 \varphi_3])),
\end{eqnarray}
\begin{eqnarray}
&& A_{12}^{*}A_{22}=\exp(2 i \varphi_2) ((\eta_2^2 + \xi_2^2) (\eta_3 \xi_1 - \eta_1
\xi_3) \cos[\varphi_3] -
   i (\eta_2^2 (\eta_3 \xi_1 + \eta_1 \xi_3) \nonumber\\&&- \xi_2^2 (\eta_3
\xi_1 + \eta_1 \xi_3) +
      2 \eta_2 \xi_2 (-\eta_1 \eta_3 + \xi_1 \xi_3))
\sin[\varphi_3]) ((\eta_2^2 + \xi_2^2) (\eta_1 \eta_3 \nonumber\\&&+ \xi_1
\xi_3) \cos[\varphi_3] +
   i (\eta_1 (\eta_2^2 \eta_3 - \eta_3 \xi_2^2 +
         2 \eta_2 \xi_2 \xi_3) + \xi_1 (2 \eta_2 \eta_3 \xi_2
- \eta_2^2 \xi_3 \nonumber\\&&+ \xi_2^2 \xi_3)) \sin[\varphi_3]),
\end{eqnarray}
\begin{eqnarray}
&& A_{22}^{*}A_{12}=\exp(-2 i \varphi_2) ((\eta_2 \xi_1 - \eta_1 \xi_2) (\eta_1 \eta_2
+ \xi_1 \xi_2) (\xi_2 (-\eta_3 + \xi_3) \nonumber\\&&+ \eta_2 (\eta_3 +
\xi_3)) (\eta_2 (\eta_3 - \xi_3) + \xi_2 (\eta_3 + \xi_3))+ (
\eta_3 \xi_2 - \eta_2 \xi_3) (\eta_2 \eta_3\nonumber\\&& + \xi_2 \xi_3) ((
\xi_1 (-\eta_2 + \xi_2) + \eta_1 (\eta_2 + \xi_2)) (\eta_1 (
\eta_2 - \xi_2) + \xi_1 (\eta_2 + \xi_2)) \cos[2 \varphi_3] \nonumber\\&&-
      i (\eta_1^2 + \xi_1^2) (\eta_2^2 + \xi_2^2) \sin[2
      \varphi_3])).\nonumber\\
\end{eqnarray}
The normalization equation takes the form:
\begin{eqnarray}
&&\langle\psi(t=TN)\vert\psi(t=TN)\rangle=(\eta_1^2 + \xi_1^2)^2 (\eta_2^2 + \xi_2^2)^2 (\eta_3^2 +
\xi_3^2).\nonumber\\
\end{eqnarray}
The normalization condition holds
\begin{eqnarray}
&&\langle\psi(t=TN)\vert\psi(t=TN)\rangle=1.
\end{eqnarray}
\section{Expectation value of $\langle\sigma_{\alpha}\rangle$, $\alpha=x,y,z$ }
\label{spindynamics}
The analytical expressions for expectation values of the spin components used in calculations:
\begin{eqnarray}\label{expectation valuesx}
&&\langle\sigma_x\rangle=\vert A_{11} \vert^2 \vert \eta_{1} \vert^2(\eta_{N}^{*}\xi_{N}+\xi_{N}^{*}\eta_{N})+A_{11}^{*}A_{12}\vert \eta_{1} \vert^2(-\vert \eta_{N}\vert^{2}+\vert \xi_{N}\vert^2)\nonumber\\&&+A_{11}^{*}A_{21}\eta_{1}^{*}\xi_{1}\exp(2i\varphi_{1})(\eta_{N}^{*}\xi_{N}+\eta_{N}\xi_{N}^{*})+ A_{11}^{*}A_{22}\eta_{1}^{*}\xi_{1}\exp(2i\varphi_{1})(-\vert \eta_{N}\vert^{2}\nonumber\\&&+\vert \xi_{N}\vert^2)+ A_{12}^{*}A_{11}\vert \eta_{1}\vert^{2}(-\vert \eta_{N}\vert^{2}+\vert \xi_{N}\vert^2)+\vert A_{12} \vert^2
\vert \eta_{1} \vert^2(-\eta_{N}^{*}\xi_{N}-\xi_{N}^{*}\eta_{N})\nonumber\\&&+ A_{12}^{*}A_{21} \eta_{1}^{*}\xi_{1}\exp(2i\varphi_{1})(-\vert \eta_{N}\vert^{2}+\vert \xi_{N}\vert^2)+A_{12}^{*}A_{22} \eta_{1}^{*}\xi_{1}\exp(2i\varphi_{1})(- \eta_{N}^{*}\xi_{N}\nonumber\\&&-\eta_{N} \xi_{N}^{*})+A_{21}^{*}A_{11} \eta_{1}\xi_{1}^{*}\exp(-2i\varphi_{1})( \eta_{N}^{*}\xi_{N}+ \xi_{N}^{*}\eta_{N})+A_{21}^{*}A_{12} \eta_{1}\xi_{1}^{*}\exp(-2i\varphi_{1})( -\vert \eta_{N}\vert^{2}\nonumber\\&&+\vert \xi_{N} \vert^{2})+\vert A_{21} \vert^2 \vert \xi_{1} \vert^2(\eta_{N}\xi_{N}^{*}+\xi_{N}\eta_{N}^{*})+A_{21}^{*}A_{22}\vert \xi_{1} \vert^2(-\vert \eta_{N}\vert^{2}+\vert \xi_{N}\vert^2)\nonumber\\&&+A_{22}^{*}A_{11} \eta_{1}\xi_{1}^{*}\exp(-2i\varphi_{1})( -\vert \eta_{N}\vert^{2}+\vert \xi_{N} \vert^{2})+A_{22}^{*}A_{12} \eta_{1}\xi_{1}^{*}\exp(-2i\varphi_{1})( - \eta_{N}^{*} \xi_{N}\nonumber\\&&-\eta_{N}\xi_{N}^{*} )+A_{21}A_{22}^{*}\vert \xi_{1}\vert^{2}( -\vert \eta_{N}\vert^{2}+\vert \xi_{N}
\vert^{2})+\vert A_{22}\vert^{2}\vert \xi\vert^{2}(-\eta_{N}\xi_{N}^{*}-\xi_{N}\eta_{N}^{*}),\nonumber\\
\end{eqnarray}
\begin{eqnarray}\label{expectation valuesy}
&&\langle\sigma_y\rangle=\vert A_{11} \vert^2 \vert \eta_{1} \vert^2(-i \eta_{N}^{*}\xi_{N}+i \xi_{N}^{*}\eta_{N})+A_{11}^{*}A_{12}\vert \eta_{1} \vert^2(i \vert \eta_{N}\vert^{2}+i\vert \xi_{N}\vert^2)\nonumber\\&&+A_{11}^{*}A_{21}\eta_{1}^{*}\xi_{1}\exp(2i\varphi_{1})(-i \eta_{N}^{*}\xi_{N}+i \eta_{N}\xi_{N}^{*})+ A_{11}^{*}A_{22}\eta_{1}^{*}\xi_{1}\exp(2i\varphi_{1})(i \vert \eta_{N}\vert^{2}\nonumber\\&&+i \vert \xi_{N}\vert^2)+ A_{12}^{*}A_{11}\vert \eta_{1}\vert^{2}(-i \vert \eta_{N}\vert^{2}-i \vert \xi_{N}\vert^2)+\vert A_{12} \vert^2 \vert \eta_{1} \vert^2(-i \eta_{N}^{*}\xi_{N}+i \xi_{N}^{*}\eta_{N})\nonumber\\&&+ A_{12}^{*}A_{21} \eta_{1}^{*}\xi_{1}\exp(2i\varphi_{1})(-i \vert \eta_{N}\vert^{2}-i \vert \xi_{N}\vert^2)+A_{12}^{*}A_{22} \eta_{1}^{*}\xi_{1}\exp(2i\varphi_{1})(-i  \eta_{N}^{*}\xi_{N}+i \eta_{N} \xi_{N}^{*})\nonumber\\&&+A_{21}^{*}A_{11} \eta_{1}\xi_{1}^{*}\exp(-2i\varphi_{1})(-i \eta_{N}^{*}\xi_{N}+i \xi_{N}^{*}\eta_{N})+A_{21}^{*}A_{12} \eta_{1}\xi_{1}^{*}\exp(-2i\varphi_{1})( i \vert \eta_{N}\vert^{2}+i \vert \xi_{N} \vert^{2})\nonumber\\&&+\vert A_{21} \vert^2 \vert \xi_{1} \vert^2(i \eta_{N}\xi_{N}^{*}-i \xi_{N}\eta_{N}^{*})+A_{21}^{*}A_{22}\vert \xi_{1} \vert^2(i \vert \eta_{N}\vert^{2}+i \vert
\xi_{N}\vert^2)\nonumber\\&&+A_{22}^{*}A_{11} \eta_{1}\xi_{1}^{*}\exp(-2i\varphi_{1})( -i \vert \eta_{N}\vert^{2}-i \vert \xi_{N} \vert^{2})+A_{22}^{*}A_{12} \eta_{1}\xi_{1}^{*}\exp(-2i\varphi_{1})( -i \eta_{N}^{*}
\xi_{N}+i \eta_{N}\xi_{N}^{*} )\nonumber\\&&+A_{21}A_{22}^{*}\vert \xi_{1}\vert^{2}( -i \vert \eta_{N}\vert^{2}-i \vert \xi_{N} \vert^{2})+\vert A_{22}\vert^{2}\vert \xi
\vert^{2}(i \eta_{N}\xi_{N}^{*}-i \xi_{N}\eta_{N}^{*}),\nonumber\\
\end{eqnarray}
\begin{eqnarray}\label{expectation valuesz}
&&\langle\sigma_z\rangle=\vert A_{11} \vert^2 \vert \eta_{1} \vert^2(\vert \eta_{N}\vert^{2}-\vert \xi_{N}\vert^{2})+A_{11}^{*}A_{12}\vert \eta_{1} \vert^2(  \xi_{N}\eta_{N}^{*}+ \xi_{N}^{*}\eta_{N})\nonumber\\&&+A_{11}^{*}A_{21}\eta_{1}^{*}\xi_{1}\exp(2i\varphi_{1})(\vert \eta_{N}\vert^{2}-\vert \xi_{N}\vert^{2})+ A_{11}^{*}A_{22}\eta_{1}^{*}\xi_{1}\exp(2i\varphi_{1})(\xi_{N}\eta_{N}^{*}+ \xi_{N}^{*}\eta_{N})\nonumber\\&&+ A_{12}^{*}A_{11}\vert \eta_{1}\vert^{2}(\xi_{N}\eta_{N}^{*}+ \xi_{N}^{*}\eta_{N})+\vert A_{12} \vert^2 \vert \eta_{1} \vert^2(\vert \xi_{N}\vert^{2}-\vert \eta_{N}\vert^{2})\nonumber\\&&+ A_{12}^{*}A_{21} \eta_{1}^{*}\xi_{1}\exp(2i\varphi_{1})( \xi_{N}^{*}\eta_{N}+\eta_{N}^{*} \xi_{N})+A_{12}^{*}A_{22} \eta_{1}^{*}\xi_{1}\exp(2i\varphi_{1})(\vert \xi_{N}\vert^{2}-\vert
\eta_{N}\vert^{2})\nonumber\\&&+A_{21}^{*}A_{11} \eta_{1}\xi_{1}^{*}\exp(-2i\varphi_{1})(\vert \eta_{N}\vert^{2}-\vert \xi_{N}\vert^{2})+A_{21}^{*}A_{12} \eta_{1}\xi_{1}^{*}\exp(-2i\varphi_{1})( \eta_{N}^{*}\xi_{N}+\xi_{N} \eta_{N})\nonumber\\&&+\vert A_{21} \vert^2 \vert \xi_{1} \vert^2(\vert
\eta_{N}\vert^{2}-\vert\xi_{N}\vert^{2})+A_{21}^{*}A_{22}\vert \xi_{1} \vert^2( \eta_{N}^{*}\xi_{N}+ \xi_{N}^{*}\eta_{N})\nonumber\\&&+A_{22}^{*}A_{11} \eta_{1}\xi_{1}^{*}\exp(-2i\varphi_{1})( \eta_{N}\xi_{N}^{*}+ \eta_{N}^{*}\xi_{N})+A_{22}^{*}A_{12} \eta_{1}\xi_{1}^{*}\exp(-2i\varphi_{1})(  \vert\xi_{N}\vert^2
-\vert \eta_{N}\vert)\nonumber\\&&+A_{21}A_{22}^{*}\vert \xi_{1}\vert^{2}( \xi_{N}^{*}\eta_{N}+ \eta_{N}^{*}\xi_{N} )+\vert A_{22}\vert^{2}\vert \xi
\vert^{2}(\vert\xi_{N}\vert^{2}-\vert\eta_{N}\vert^{2}).\nonumber\\
\end{eqnarray}
\section{Elements of density matrix for Eq.({\ref{densityformalism}})}
\label{Quantumcoherence}
The elements of the reduced density matrix, analytical expressions used for calculation
of the coherence.
\begin{eqnarray}
&&\rho_{11}=\vert A_{11} \vert^2\vert \eta_{1} \vert^2\vert \eta_{N} \vert^2+A_{11}^{*}A_{12}\vert \eta_{1} \vert^2\eta_{N}^{*}\xi_{N}+A_{11}^{*}A_{21}\eta_{1}^{*}\xi_{1}\exp{(2 i \varphi_{1})}\vert \eta_{N}\vert^2\nonumber\\&&+A_{11}^{*}A_{22}\eta_{1}^{*}\xi_{1}\exp{(2 i \varphi_{1})}\eta_{N}^{*}\xi_{N}+A_{12}^{*}A_{11}\vert \eta_{1} \vert^2\eta_{N}\xi_{N}^{*}+\vert A_{12} \vert^2\vert \eta_{1} \vert^2\vert \xi_{N} \vert^2\nonumber\\&&+A_{12}^{*}A_{21}\eta_{1}^{*}\xi_{1}\exp{(2 i \varphi_{1})} \eta_{N}\xi_{N}^{*}+A_{12}^{*}A_{22}\eta_{1}^{*}\xi_{1}\exp{(2 i \varphi_{1})} \vert\xi_{N}\vert^2\nonumber\\&&+A_{21}^{*}A_{11} \eta_{1} \xi_{1}^{*}\exp{(-2 i \varphi_{1})}\vert\eta_{N}\vert^2+A_{21}^{*}A_{12}\eta_{1}\xi_{1}^{*}\exp{(-2 i \varphi_{1})}\eta_{N}^{*}\xi_{N}\nonumber\\&&+\vert A_{21}\vert^2\vert \xi_{1} \vert^{2} \vert\eta_{N}\vert^2+A_{21}^{*}A_{22}\vert\xi_{1}\vert^2\xi_{N}\eta_{N}^{*}+A_{11}A_{22}^{*}\exp{(-2 i \varphi_{1})}\eta_{N}\xi_{N}^{*}\nonumber\\&&+A_{12}A_{22}^{*}\eta_{1}\xi_{1}^{*}\exp{(-2 i \varphi_{1})}\vert\xi_{N}\vert^2+A_{21}A_{22}^{*}\vert \xi_{1} \vert^2 \eta_{N}\xi_{N}^{*}+\vert A_{22} \vert^2 \vert \xi_{1} \vert^2\vert \xi_{N} \vert^2,\nonumber\\
\end{eqnarray}\\
\begin{eqnarray}
&&\rho_{12}=\vert A_{11} \vert^2\vert \eta_{1} \vert^2   \eta_{N} \xi_{N}^{*}+A_{11}^{*}A_{12}\vert \eta_{1} \vert^2\vert\xi_{N}\vert^{2}+A_{11}^{*}A_{21}\eta_{1}^{*}\xi_{1}\exp{(2 i \varphi_{1})} \eta_{N}\xi_{N}^{*}\nonumber\\&&+A_{11}^{*}A_{22}\eta_{1}^{*}\xi_{1}\exp{(2 i \varphi_{1})}\vert \xi_{N}\vert^{2}-A_{12}^{*}A_{11}\vert \eta_{1} \vert^2\vert\eta_{N}\vert^2-\vert A_{12} \vert^2\vert \eta_{1} \vert^2\xi_{N} \eta_{N}^{*}\nonumber\\&&-A_{12}^{*}A_{21}\eta_{1}^{*}\xi_{1}\exp{(2 i \varphi_{1})} \vert \eta_{N}\vert^{2}-A_{12}^{*}A_{22}\eta_{1}^{*}\xi_{1}\exp{(2 i \varphi_{1})}\eta_{N}^{*}\xi_{N}\nonumber\\&&+A_{21}^{*}A_{11} \eta_{1} \xi_{1}^{*}\exp{(-2 i \varphi_{1})}\xi_{N}^{*}\eta_{N}+A_{21}^{*}A_{12}\eta_{1}\xi_{1}^{*}\exp{(-2 i \varphi_{1})}\vert\xi_{N}\vert^{2}\nonumber\\&&+\vert A_{21}\vert^2\vert \xi_{1} \vert^{2} \eta_{N}\xi_{N}^{*}+A_{21}^{*}A_{22}\vert\xi_{1}\vert^2\vert\xi_{N}\vert^2-A_{11}A_{22}^{*}\eta_{1}\xi_{1}^{*}\exp{(-2 i \varphi_{1})}\vert\eta_{N}\vert^2\nonumber\\&&-A_{12}A_{22}^{*}\eta_{1}\xi_{1}^{*}\exp{(-2 i \varphi_{1})}\eta_{N}^{*}\xi_{N}-A_{21}A_{22}^{*}\vert \xi_{1} \vert^2  \vert\eta_{N}\vert^{2}-\vert A_{22} \vert^2 \vert \xi_{1} \vert^2  \xi_{N} \eta_{N}^{*}, \nonumber\\
\end{eqnarray}
\\
\begin{eqnarray}
&&\rho_{21}=\vert A_{11} \vert^2\vert \eta_{1} \vert^2   \eta_{N}^{*} \xi_{N}-A_{11}^{*}A_{12}\vert \eta_{1} \vert^2\vert\eta_{N}\vert^{2}+A_{11}^{*}A_{21}\eta_{1}^{*}\xi_{1}\exp{(2 i \varphi_{1})}
\eta_{N}^{*}\xi_{N}\nonumber\\&&-A_{11}^{*}A_{22}\eta_{1}^{*}\xi_{1}\exp{(2 i \varphi_{1})}\vert \eta_{N}\vert^{2}+A_{12}^{*}A_{11}\vert \eta_{1} \vert^2\vert\xi_{N}\vert^2-\vert A_{12} \vert^2\vert \eta_{1} \vert^2\xi_{N}^{*} \eta_{N}\nonumber\\&&+A_{12}^{*}A_{21}\eta_{1}^{*}\xi_{1}\exp{(2 i \varphi_{1})} \vert \xi_{N}\vert^{2}-A_{12}^{*}A_{22}\eta_{1}^{*}\xi_{1}\exp{(2 i \varphi_{1})}\eta_{N}\xi_{N}^{*}\nonumber\\&&+A_{21}^{*}A_{11} \eta_{1} \xi_{1}^{*}\exp{(-2 i \varphi_{1})}\xi_{N}\eta_{N}^{*}-A_{21}^{*}A_{12}\eta_{1}\xi_{1}^{*}\exp{(-2 i \varphi_{1})}\vert\eta_{N}\vert^{2}\nonumber\\&&+\vert A_{21}\vert^2\vert \xi_{1} \vert^{2} \eta_{N}^{*}\xi_{N}-A_{21}^{*}A_{22}\vert\xi_{1}\vert^2\vert\eta_{N}\vert^2+A_{11}A_{22}^{*}\eta_{1}\xi_{1}^{*}\exp{(-2 i \varphi_{1})}\vert\xi_{N}\vert^2\nonumber\\&&-A_{12}A_{22}^{*}\eta_{1}\xi_{1}^{*}\exp{(-2 i \varphi_{1})}\eta_{N}\xi_{N}^{*}+A_{21}A_{22}^{*}\vert \xi_{1} \vert^2  \vert\xi_{N}\vert^{2}-\vert A_{22} \vert^2 \vert \xi_{1} \vert^2  \xi_{N}^{*} \eta_{N},\nonumber\\
\end{eqnarray}\\
\begin{eqnarray}
&&\rho_{22}=\vert A_{11} \vert^2\vert \eta_{1} \vert^2\vert \xi_{N} \vert^2-A_{11}^{*}A_{12}\vert \eta_{1} \vert^2\eta_{N}\xi_{N}^{*}+A_{11}^{*}A_{21}\eta_{1}^{*}\xi_{1}\exp{(2 i \varphi_{1})}\vert \xi_{N}\vert^2\nonumber\\&&-A_{11}^{*}A_{22}\eta_{1}^{*}\xi_{1}\exp{(2 i \varphi_{1})}\eta_{N}\xi_{N}^{*}-A_{12}^{*}A_{11}\vert
\eta_{1} \vert^2\eta_{N}^{*}\xi_{N}+\vert A_{12} \vert^2\vert \eta_{1} \vert^2\vert \eta_{N} \vert^2\nonumber\\&&-A_{12}^{*}A_{21}\eta_{1}^{*}\xi_{1}\exp{(2 i \varphi_{1})} \eta_{N}^{*}\xi_{N}+A_{12}^{*}A_{22}\eta_{1}^{*}\xi_{1}\exp{(2 i \varphi_{1})} \vert\eta_{N}\vert^2\nonumber\\&&+A_{21}^{*}A_{11} \eta_{1} \xi_{1}^{*}\exp{(-2 i \varphi_{1})}\vert\xi_{N}\vert^2-A_{21}^{*}A_{12}\eta_{1}\xi_{1}^{*}\exp{(-2 i \varphi_{1})}\eta_{N}\xi_{N}^{*}\nonumber\\&&+\vert A_{21}\vert^2\vert \xi_{1} \vert^{2} \vert\xi_{N}\vert^2-A_{21}^{*}A_{22}\vert\xi_{1}\vert^2\xi_{N}^{*}\eta_{N}^{*}-A_{11}A_{22}^{*}\exp{(-2 i \varphi_{1})}\eta_{N}^{*}\xi_{N}\nonumber\\&&+A_{12}A_{22}^{*}\eta_{1}\xi_{1}^{*}\exp{(-2 i \varphi_{1})}\vert\eta_{N}\vert^2-A_{21}A_{22}^{*}\vert \xi_{1} \vert^2 \eta_{N}^{*}\xi_{N}+\vert A_{22} \vert^2 \vert \xi_{1} \vert^2\vert \eta_{N} \vert^2.\nonumber\\
\end{eqnarray}
\section{Normalization constants for eigenstates of Floquet operator for Eq.(\ref{spectral decomposition spin-1})   and matrix elements for Eq.(\ref{time dependent spin-1x})}
\label{wavefunctionforspin-1}
The eigenstates of Floquet operator for spin-1  is defined as:\nonumber\\
 $\vert\varphi_{n}^1\rangle=-\eta_n\vert0\rangle+\xi_n \vert1\rangle+\zeta_n \vert2\rangle,$ and 
 $\vert\varphi_{n}^2\rangle=x_n\vert0\rangle+y_n \vert1\rangle+z_n \vert2\rangle$and
 $\vert\varphi_{n}^3\rangle=u_n\vert0\rangle+v_n \vert1\rangle+w_n \vert2\rangle$, 
where the normalization constants for above eigenstates is defined  as:
 $\eta_n=-\frac{1}{\sqrt{2}}=-\zeta_n $,
 $\xi_n=0$, $x_{n}=\frac{a_{n}}{\sqrt{a_{n}^2+b_{n}^2+c_{n}^2}}$,$y_{n}=\frac{b_{n}}{\sqrt{a_{n}^2+b_{n}^2+c_{n}^2}}$,$z_{n}=\frac{c_{n}}{\sqrt{a_{n}^2+b_{n}^2+c_{n}^2}}$,
 $u_{n}=\frac{d_{n}}{\sqrt{d_{n}^2+e_{n}^2+f_{n}^2}}$,$v_{n}=\frac{e_{n}}{\sqrt{d_{n}^2+e_{n}^2+f_{n}^2}}$,$w_{n}=\frac{f_{n}}{\sqrt{d_{n}^2+e_{n}^2+f_{n}^2}}$,
 $a_{n}=\frac{2\chi_{n}^{2}+2\sqrt{2}\chi_{n}\Omega+\Omega^{2}}{(\sqrt{2}\chi_{n}+\Omega)^2}$, $b_{n}=\frac{\delta-\sqrt{4 \chi_{n}^2+\delta^2+4\sqrt{2}\chi_{n}\Omega+2 \Omega^2}}{\sqrt{2}\chi_{n}+\Omega}$, $c_{n}=1$,
 $d_{n}=\frac{2\chi_{n}^{2}+2\sqrt{2}\chi_{n}\Omega+\Omega^{2}}{(\sqrt{2}\chi_{n}+\Omega)^2}$, $e_{n}=\frac{\delta+\sqrt{4 \chi_{n}^2+\delta^2+4\sqrt{2}\chi_{n}\Omega+2 \Omega^2}}{\sqrt{2}\chi_{n}+\Omega}$, $f_{n}=1$.\\
Matrix elements of Eq.\ref{time dependent spin-1x} are given as. 
\begin{eqnarray}
 G_{11}&=&\exp(-i\varphi_{n}^{1})(\eta_{n}\eta_{n-1}+\xi_{n}\xi_{n-1}+\eta_{n}\eta_{n-1}),
\end{eqnarray}  
\begin{eqnarray}
 G_{12}&=&\exp(-i\varphi_{n}^{1})(-\eta_{n}x_{n-1}+\xi_{n}y_{n-1}+\eta_{n}z_{n-1}),
\end{eqnarray}
\begin{eqnarray}
 G_{13}&=& \exp(-i\varphi_{n}^{1})(-\eta_{n}u_{n-1}+\xi_{n}v_{n-1}+\eta_{n}w_{n-1}),
\end{eqnarray}
\begin{eqnarray}
 G_{21}&= &\exp(-i\varphi_{n}^{2})(-x_{n}\eta_{n-1}+y_{n}\xi_{n-1}+z_{n}\eta_{n-1}),
\end{eqnarray}
\begin{eqnarray}
 G_{22}&=& \exp(i\varphi_{n}^{2})(x_{n}x_{n-1}+y_{n}y_{n-1}+z_{n}z_{n-1}),
\end{eqnarray}
\begin{eqnarray}
 G_{23}&=& \exp(i\varphi_{n}^{2})(x_{n}u_{n-1}+y_{n}v_{n-1}+z_{n}w_{n-1}),
\end{eqnarray}
\begin{eqnarray}
 G_{31} &=& \exp(-i\varphi_{n}^{3})(-u_{n}\eta_{n-1}+v_{n}\xi_{n-1}+w_{n}\eta_{n-1}),
\end{eqnarray}
\begin{eqnarray}
 G_{32}&=& \exp(i\varphi_{n}^{2})(u_{n}x_{n-1}+v_{n}y_{n-1}+w_{n}z_{n-1}),
\end{eqnarray}
\begin{eqnarray}
 G_{33} &=&  \exp(i\varphi_{n}^{2})(u_{n}u_{n-1}+v_{n}v_{n-1}+w_{n}w_{n-1}).
\end{eqnarray}

 \bibliographystyle{elsarticle-num} 
 \bibliography{main}

\begin{thebibliography}{10}
\expandafter\ifx\csname url\endcsname\relax
  \def\url#1{\texttt{#1}}\fi
\expandafter\ifx\csname urlprefix\endcsname\relax\def\urlprefix{URL }\fi
\expandafter\ifx\csname href\endcsname\relax
  \def\href#1#2{#2} \def\path#1{#1}\fi

\bibitem{el2007deterministic}
M.~El~Naschie, Deterministic quantum mechanics versus classical mechanical
  indeterminism, International Journal of Nonlinear Sciences and Numerical
  Simulation 8~(1) (2007) 5--10.

\bibitem{PhysRevA.100.062107}
F.~Del~Santo, N.~Gisin, Physics without determinism: Alternative
  interpretations of classical physics, Phys. Rev. A 100 (2019) 062107.

\bibitem{cresson2019selection}
J.~Cresson, Y.~Kheloufi, K.~Nachi, F.~Pierret, Selection of a stochastic
  landau-lifshitz equation and the stochastic persistence problem, Journal of
  Mathematical Physics 60~(8) (2019) 083512.

\bibitem{chirikov1979universal}
B.~V. Chirikov, et~al., A universal instability of many-dimensional oscillator
  systems, Physics reports 52~(5) (1979) 263--379.

\bibitem{casati1979stochastic}
G.~Casati, B.~Chirikov, F.~Izraelev, J.~Ford, Stochastic behavior of a quantum
  pendulum under a periodic perturbation, in: Stochastic behavior in classical
  and quantum Hamiltonian systems, Springer, 1979, pp. 334--352.

\bibitem{lichtenberg2013regular}
A.~J. Lichtenberg, M.~A. Lieberman, Regular and chaotic dynamics, Vol.~38,
  Springer Science \& Business Media, 2013.

\bibitem{moore1995atom}
F.~Moore, J.~Robinson, C.~Bharucha, B.~Sundaram, M.~Raizen, Atom optics
  realization of the quantum $\delta$-kicked rotor, Physical Review Letters
  75~(25) (1995) 4598.

\bibitem{PhysRevLett.87.074102}
M.~B. d'Arcy, R.~M. Godun, M.~K. Oberthaler, D.~Cassettari, G.~S. Summy,
  Quantum enhancement of momentum diffusion in the delta-kicked rotor, Phys.
  Rev. Lett. 87 (2001) 074102.
\newblock \href {https://doi.org/10.1103/PhysRevLett.87.074102}
  {\path{doi:10.1103/PhysRevLett.87.074102}}.

\bibitem{Tanaka_1996}
A.~Tanaka, Quantum mechanical entanglements with chaotic dynamics, Journal of
  Physics A: Mathematical and General 29~(17) (1996) 5475--5497.
\newblock \href {https://doi.org/10.1088/0305-4470/29/17/020}
  {\path{doi:10.1088/0305-4470/29/17/020}}.

\bibitem{casati1985energy}
G.~Casati, B.~Chirikov, I.~Guarneri, Energy-level statistics of integrable
  quantum systems, Physical review letters 54~(13) (1985) 1350.

\bibitem{shepelyansky1987localization}
D.~Shepelyansky, Localization of diffusive excitation in multi-level systems,
  Physica D: Nonlinear Phenomena 28~(1-2) (1987) 103--114.

\bibitem{PhysRevA.44.4704}
G.~J. Milburn, C.~A. Holmes, Quantum coherence and classical chaos in a pulsed
  parametric oscillator with a kerr nonlinearity, Phys. Rev. A 44 (1991)
  4704--4711.
\newblock \href {https://doi.org/10.1103/PhysRevA.44.4704}
  {\path{doi:10.1103/PhysRevA.44.4704}}.

\bibitem{Naik2009}
A.~K. Naik, M.~S. Hanay, W.~K. Hiebert, X.~L. Feng, M.~L. Roukes, {Towards
  single-molecule nanomechanical mass spectrometry}, Nature Nanotechnology
  4~(7) (2009) 445.
\newblock \href {https://doi.org/10.1038/nnano.2009.152}
  {\path{doi:10.1038/nnano.2009.152}}.

\bibitem{Connell2010}
A.~D.~O. Connell, M.~Hofheinz, M.~Ansmann, R.~C. Bialczak, M.~Lenander,
  E.~Lucero, M.~Neeley, D.~Sank, H.~Wang, M.~Weides, J.~Wenner, J.~M. Martinis,
  A.~N. Cleland, {Quantum ground state and single-phonon control of a
  mechanical resonator}, Nature 464~(7289) (2010) 697.
\newblock \href {https://doi.org/10.1038/nature08967}
  {\path{doi:10.1038/nature08967}}.

\bibitem{Alegre2011}
T.~P.~M. Alegre, J.~Chan, M.~Eichenfield, M.~Winger, Q.~Lin, J.~T. Hill, D.~E.
  Chang, O.~Painter, Electromagnetically induced transparency and slow light
  with optomechanics, Nature 472~(7341) (2011) 69.
\newblock \href {https://doi.org/10.1038/nature09933}
  {\path{doi:10.1038/nature09933}}.

\bibitem{Stannigel2010}
K.~Stannigel, P.~Rabl, A.~S. S{\o}rensen, P.~Zoller, M.~D. Lukin,
  {Optomechanical transducers for long-distance quantum communication}, Phys.
  Rev. Lett. 105~(22) (2010) 220501.
\newblock \href {https://doi.org/10.1103/PhysRevLett.105.220501}
  {\path{doi:10.1103/PhysRevLett.105.220501}}.

\bibitem{Safavi-Naeini2011}
A.~H. Safavi-Naeini, O.~Painter, Proposal for an optomechanical traveling wave
  phonon--photon translator, New Journal of Physics 13~(1) (2011) 013017.

\bibitem{Camerer2011}
S.~Camerer, M.~Korppi, A.~J{\"{o}}ckel, D.~Hunger, T.~W. H{\"{a}}nsch,
  P.~Treutlein, {Realization of an optomechanical interface between ultracold
  atoms and a membrane}, Phys. Rev. Lett. 107~(22) (2011) 223001.
\newblock \href {https://doi.org/10.1103/PhysRevLett.107.223001}
  {\path{doi:10.1103/PhysRevLett.107.223001}}.

\bibitem{Eichenfield2009}
M.~Eichenfield, J.~Chan, R.~M. Camacho, K.~J. Vahala, O.~Painter,
  {Optomechanical crystals}, Nature 462~(7269) (2009) 78--82.
\newblock \href {https://doi.org/10.1038/nature08524}
  {\path{doi:10.1038/nature08524}}.

\bibitem{Safavi-Naeini2012}
A.~H. Safavi-Naeini, J.~Chan, J.~T. Hill, T.~P.~M. Alegre, A.~Krause,
  O.~Painter, {Observation of quantum motion of a nanomechanical resonator},
  Phys. Rev. Lett. 108~(3) (2012) 033602.
\newblock \href {https://doi.org/10.1103/PhysRevLett.108.033602}
  {\path{doi:10.1103/PhysRevLett.108.033602}}.

\bibitem{Brahms2012}
N.~Brahms, T.~Botter, S.~Schreppler, D.~W.~C. Brooks, D.~M. Stamper-Kurn,
  {Optical detection of the quantization of collective atomic motion}, Phys.
  Rev. Lett. 108~(13) (2012) 133601.
\newblock \href {https://doi.org/10.1103/PhysRevLett.108.133601}
  {\path{doi:10.1103/PhysRevLett.108.133601}}.

\bibitem{Nunnenkamp2012}
A.~Nunnenkamp, K.~B{\o}rkje, S.~M. Girvin, {Cooling in the single-photon
  strong-coupling regime of cavity optomechanics}, Phys. Rev. A 85~(5) (2012)
  051803(R).
\newblock \href {https://doi.org/10.1103/PhysRevA.85.051803}
  {\path{doi:10.1103/PhysRevA.85.051803}}.

\bibitem{Khalili2012}
F.~Y. Khalili, H.~Miao, H.~Yang, A.~H. Safavi-Naeini, O.~Painter, Y.~Chen,
  {Quantum back-action in measurements of zero-point mechanical oscillations},
  Phys. Rev. A 86~(3) (2012) 033602.
\newblock \href {https://doi.org/10.1103/PhysRevA.86.033840}
  {\path{doi:10.1103/PhysRevA.86.033840}}.

\bibitem{Meaney2011}
C.~P. Meaney, R.~H. McKenzie, G.~J. Milburn, {Quantum entanglement between a
  nonlinear nanomechanical resonator and a microwave field}, Phys. Rev. E
  83~(5) (2011) 056202.
\newblock \href {https://doi.org/10.1103/PhysRevE.83.056202}
  {\path{doi:10.1103/PhysRevE.83.056202}}.

\bibitem{Atalaya2011}
J.~Atalaya, A.~Isacsson, M.~I. Dykman, {Diffusion-induced dephasing in
  nanomechanical resonators}, Phys. Rev. B 83~(4) (2011) 045419.
\newblock \href {https://doi.org/10.1103/PhysRevB.83.045419}
  {\path{doi:10.1103/PhysRevB.83.045419}}.

\bibitem{Rabl2010}
P.~Rabl, {Cooling of mechanical motion with a two-level system: The
  high-temperature regime}, Phys. Rev. B 82~(16) (2010) 165320.
\newblock \href {https://doi.org/10.1103/PhysRevB.82.165320}
  {\path{doi:10.1103/PhysRevB.82.165320}}.

\bibitem{Prants2011}
S.~V. Prants, {A group-theoretical approach to study atomic motion in a laser
  field}, J. Phys. A 44~(26) (2011) 265101.
\newblock \href {https://doi.org/10.1088/1751-8113/44/26/265101}
  {\path{doi:10.1088/1751-8113/44/26/265101}}.

\bibitem{Ludwig2010}
M.~Ludwig, K.~Hammerer, F.~Marquardt, {Entanglement of mechanical oscillators
  coupled to a nonequilibrium environment}, Phys. Rev. A 82~(1) (2010) 012333.
\newblock \href {https://doi.org/10.1103/PhysRevA.82.012333}
  {\path{doi:10.1103/PhysRevA.82.012333}}.

\bibitem{Schmidt2010}
T.~L. Schmidt, K.~B{\o}rkje, C.~Bruder, B.~Trauzettel, {Detection of
  qubit-oscillator entanglement in nanoelectromechanical systems}, Phys. Rev.
  Lett. 104~(17) (2010) 177205.
\newblock \href {https://doi.org/10.1103/PhysRevLett.104.177205}
  {\path{doi:10.1103/PhysRevLett.104.177205}}.

\bibitem{Karabalin2009}
R.~B. Karabalin, M.~C. Cross, M.~L. Roukes, {Nonlinear dynamics and chaos in
  two coupled nanomechanical resonators}, Phys. Rev. B 79~(16) (2009) 165309.
\newblock \href {https://doi.org/10.1103/PhysRevB.79.165309}
  {\path{doi:10.1103/PhysRevB.79.165309}}.

\bibitem{Chotorlishvili_2011}
L.~Chotorlishvili, A.~Ugulava, G.~Mchedlishvili, A.~Komnik, S.~Wimberger,
  J.~Berakdar, {Nonlinear dynamics of two coupled nano-electromechanical
  resonators}, J. Phys. B 44~(21) (2011) 215402.
\newblock \href {https://doi.org/10.1088/0953-4075/44/21/215402}
  {\path{doi:10.1088/0953-4075/44/21/215402}}.

\bibitem{Shevchenko2012}
S.~N. Shevchenko, A.~N. Omelyanchouk, E.~Il'ichev, {Multiphoton transitions in
  Josephson-junction qubits (Review Article)}, Low Temperature Physics 38~(4)
  (2012) 283--300.
\newblock \href {https://doi.org/10.1063/1.3701717}
  {\path{doi:10.1063/1.3701717}}.

\bibitem{Liu2010}
Y.~X. Liu, A.~Miranowicz, Y.~B. Gao, J.~Bajer, C.~P. Sun, F.~Nori,
  {Qubit-induced phonon blockade as a signature of quantum behavior in
  nanomechanical resonators}, Phys. Rev. A 82~(3) (2010) 032101.
\newblock \href {https://doi.org/10.1103/PhysRevA.82.032101}
  {\path{doi:10.1103/PhysRevA.82.032101}}.

\bibitem{Shevchenko2010}
S.~Shevchenko, S.~Ashhab, F.~Nori, {Landau–Zener–St{\"{u}}ckelberg
  interferometry}, Physics Reports 492~(1) (2010) 1--30.
\newblock \href {https://doi.org/10.1016/j.physrep.2010.03.002}
  {\path{doi:10.1016/j.physrep.2010.03.002}}.

\bibitem{Zueco2009}
D.~Zueco, G.~M. Reuther, S.~Kohler, P.~H{\"{a}}nggi, {Qubit-oscillator dynamics
  in the dispersive regime: Analytical theory beyond the rotating-wave
  approximation}, Phys. Rev. A 80~(3) (2009) 033846.
\newblock \href {https://doi.org/10.1103/PhysRevA.80.033846}
  {\path{doi:10.1103/PhysRevA.80.033846}}.

\bibitem{Cohen2013}
G.~Z. Cohen, M.~{Di Ventra}, {Reading, writing, and squeezing the entangled
  states of two nanomechanical resonators coupled to a SQUID}, Phys. Rev. B
  87~(1) (2013) 014513.
\newblock \href {https://doi.org/10.1103/PhysRevB.87.014513}
  {\path{doi:10.1103/PhysRevB.87.014513}}.

\bibitem{Rabl2009}
P.~Rabl, P.~Cappellaro, M.~V.~G. Dutt, L.~Jiang, J.~R. Maze, M.~D. Lukin,
  Strong magnetic coupling between an electronic spin qubit and a mechanical
  resonator, Phys. Rev. B 79 (2009) 041302.
\newblock \href {https://doi.org/10.1103/PhysRevB.79.041302}
  {\path{doi:10.1103/PhysRevB.79.041302}}.

\bibitem{Zhou2010}
L.~G. Zhou, L.~F. Wei, M.~Gao, X.~B. Wang, {Strong coupling between two distant
  electronic spins via a nanomechanical resonator}, Phys. Rev. A 81~(4) (2010)
  042323.
\newblock \href {https://doi.org/10.1103/PhysRevA.81.042323}
  {\path{doi:10.1103/PhysRevA.81.042323}}.

\bibitem{Chotorlishvili2013}
L.~Chotorlishvili, D.~Sander, A.~Sukhov, V.~Dugaev, V.~R. Vieira, A.~Komnik,
  J.~Berakdar, {Entanglement between nitrogen vacancy spins in diamond
  controlled by a nanomechanical resonator}, Phys. Rev. B 88~(8) (2013) 085201.
\newblock \href {https://doi.org/10.1103/PhysRevB.88.085201}
  {\path{doi:10.1103/PhysRevB.88.085201}}.

\bibitem{Toklikishvili2017}
Z.~Toklikishvili, L.~Chotorlishvili, S.~K. Mishra, S.~Stagraczynski,
  M.~Sch{\"{u}}ler, A.~R.~P. Rau, J.~Berakdar, Entanglement dynamics of two
  nitrogen vacancy centers coupled by a nanomechanical resonator, J. Phys. B
  50~(5) (2017) 055007.
\newblock \href {https://doi.org/10.1088/1361-6455/aa5a69}
  {\path{doi:10.1088/1361-6455/aa5a69}}.

\bibitem{Chotorlishvili2010}
L.~Chotorlishvili, A.~Ugulava, {Quantum chaos and its kinetic stage of
  evolution}, Physica D: Nonlinear Phenomena 239~(3-4) (2010) 103--122.
\newblock \href {https://doi.org/10.1016/j.physd.2009.08.017}
  {\path{doi:10.1016/j.physd.2009.08.017}}.

\bibitem{zaslavsky2007physics}
G.~M. Zaslavsky, The physics of chaos in Hamiltonian systems, world scientific,
  2007.

\bibitem{Chotorlishvili2018}
L.~Chotorlishvili, P.~Zi{\k{e}}ba, I.~Tralle, A.~Ugulava, {Zitterbewegung and
  symmetry switching in Klein's four-group}, J. Phys. A 51~(3) (2018) 035004.
\newblock \href {https://doi.org/10.1088/1751-8121/aa9a2c}
  {\path{doi:10.1088/1751-8121/aa9a2c}}.

\bibitem{ugulava2005irreversible}
A.~Ugulava, L.~Chotorlishvili, K.~Nickoladze, Irreversible evolution of quantum
  chaos, Phys. Rev. E 71 (2005) 056211.
\newblock \href {https://doi.org/10.1103/PhysRevE.71.056211}
  {\path{doi:10.1103/PhysRevE.71.056211}}.

\bibitem{haake1991quantum}
F.~Haake, Quantum signatures of chaos, in: Quantum Coherence in Mesoscopic
  Systems, Springer, 1991, pp. 583--595.

\bibitem{khomitsky2013spin}
D.~Khomitsky, A.~Malyshev, E.~Y. Sherman, M.~Di~Ventra, Spin chaos
  manifestation in a driven quantum billiard with spin-orbit coupling, Physical
  Review B 88~(19) (2013) 195407.

\bibitem{RevModPhys.89.041003}
A.~Streltsov, G.~Adesso, M.~B. Plenio, Colloquium: Quantum coherence as a
  resource, Reviews of Modern Physics 89~(4) (2017) 041003.

\bibitem{PhysRevLett.115.020403}
A.~Streltsov, U.~Singh, H.~S. Dhar, M.~N. Bera, G.~Adesso, Measuring quantum
  coherence with entanglement, Phys. Rev. Lett. 115 (2015) 020403.
\newblock \href {https://doi.org/10.1103/PhysRevLett.115.020403}
  {\path{doi:10.1103/PhysRevLett.115.020403}}.

\bibitem{zou2012phase}
Y.~Zou, R.~V. Donner, M.~Wickramasinghe, I.~Z. Kiss, M.~Small, J.~Kurths, Phase
  coherence and attractor geometry of chaotic electrochemical oscillators,
  Chaos: An Interdisciplinary Journal of Nonlinear Science 22~(3) (2012)
  033130.

\bibitem{bocchieri1957quantum}
P.~Bocchieri, A.~Loinger, Quantum recurrence theorem, Physical Review 107~(2)
  (1957) 337.

\bibitem{PhysRevLett.48.711}
T.~Hogg, B.~A. Huberman, Recurrence phenomena in quantum dynamics, Phys. Rev.
  Lett. 48 (1982) 711--714.
\newblock \href {https://doi.org/10.1103/PhysRevLett.48.711}
  {\path{doi:10.1103/PhysRevLett.48.711}}.

\bibitem{casati1999quantum}
G.~Casati, G.~Maspero, D.~L. Shepelyansky, Quantum poincar{\'e} recurrences,
  Physical review letters 82~(3) (1999) 524.

\bibitem{PhysRevLett.109.160401}
F.~Wilczek, Quantum time crystals, Phys. Rev. Lett. 109 (2012) 160401.
\newblock \href {https://doi.org/10.1103/PhysRevLett.109.160401}
  {\path{doi:10.1103/PhysRevLett.109.160401}}.

\bibitem{PhysRevLett.117.090402}
D.~V. Else, B.~Bauer, C.~Nayak, Floquet time crystals, Phys. Rev. Lett. 117
  (2016) 090402.
\newblock \href {https://doi.org/10.1103/PhysRevLett.117.090402}
  {\path{doi:10.1103/PhysRevLett.117.090402}}.

\bibitem{khomitsky2016regular}
D.~Khomitsky, A.~Chubanov, A.~Konakov, Regular and irregular dynamics of
  spin-polarized wavepackets in a mesoscopic quantum dot at the edge of
  topological insulator, Journal of Experimental and Theoretical Physics
  123~(6) (2016) 1043--1059.

\end{thebibliography}





\end{document}